\documentclass[a4paper,11pt]{article}
\usepackage{jinstpub} 
\usepackage{lineno}
\usepackage{here}
\usepackage{subcaption}
\usepackage[usenames,dvipsnames]{color}
\usepackage{multicol}

\title{Preliminary Results of the 2023 International Fermilab Booster Studies}

\author[a]{J. Eldred,\note{Corresponding author.}}
\author[a]{M.~Balcewicz,}
\author[b]{F.~Schmidt,}
\author[c]{and B.~Simons}
\affiliation[a]{Fermi National Accelerator Laboratory, \\ Batavia, Illinois 60510 U.S.A.}
\affiliation[b]{European Organization for Nuclear Research,\\ CERN CH-1211, Gen\`{e}ve 23, Switzerland}
\affiliation[c]{Department of Physics, Northern Illinois University,\\ DeKalb, IL, 60115, U.S.A.}

\emailAdd{jseldred@fnal.gov}

\abstract{An overview is given of the methods and preliminary results from dedicated beam studies on three topics conducted over five days in July 2023. In the first study, the Fermilab Booster magnets were held constant at magnetic fields corresponding to the injection energy. The beam loss and emittance growth were observed under varying intensity, tunes, and sextupole resonances. The corresponding beam conditions were also simulated with the MADX-SC code~\cite{Schmidt:2644660}. In the second study, measurements of the vertical half-integer resonance and correction methods are conducted for high-intensity beams ramping in the Booster. Finally, syncho-betatron instabilities are observed during transition-crossing in the Booster under strong space-charge conditions.}


\begin{document}
\maketitle
\flushbottom

\section{Introduction}
\label{sec:intro}

The Fermilab Booster is a 8~GeV proton synchrotron at the heart of the Fermilab particle physics program. Every 15~Hz, the Fermilab Booster accumulates $\sim$4.5e12 proton from the 400~MeV H$^{-}$ linac and accelerates those protons to 8~GeV. Next, the 8~GeV protons are extracted to the short-baseline neutrino program~\cite{SBND}, rebunched in the Fermilab Recycler for the muon campus program~\cite{mu2e}, or accumulated in the Fermilab Recycler Ring to be further accelerated in the Main Injector program. In the Main Injector the protons are accelerated to 120~GeV and delivered to the long-baseline neutrino program. Presently the long-baseline neutrino beamline is the Neutrinos at the Main Injector (NuMI) which delivers neutrinos to the NuMI Off-Axis $\nu_{e}$ Appearance (NOvA) experiment~\cite{NOvA}. In the near future it will be the Long Baseline Neutrino Facility (LBNF) beamline and the Deep Underground Neutrino Experiment (LBNF/DUNE) program~\cite{DUNE}.

Fermilab is currently undergoing the Proton Improvement Plan II (PIP-II) accelerator upgrade in preparation for LBNF/DUNE~\cite{PIP2}. The PIP-II upgrade includes a new 2~mA CW-capable SRF linac for 800~MeV injection into the Booster, an upgrade of the Booster ramp rate from 15~Hz to 20~Hz, and increase in Booster intensity to 6.5e12 protons. The Fermilab Booster will be nearly 60~years at the beginning of PIP-II operations, and therefore ensuring reliable performance of the Booster will be critical to the success of the ambitious neutrino physics program.

The 2023 P5 report~\cite{P5} affirms support for the PIP-II and LBNF/DUNE, as well as the newly proposed Accelerator Complex Evolution Main Injector Ramp \& Targetry (ACE-MIRT) upgrade, which would increase LBNF beam power from 1.2~MW to more than 2~MW by reducing the Main Injector ramp rate, developing new high power neutrino targets, and modernizing accelerator infrastructure to guarantee reliable operations. In Area Recommendation 13, the P5 report specifically recommends the community ``Assess the booster synchrotron and related systems for reliability risks through the first decade of DUNE operation, and take measures to preemptively address these risks.'' Therefore, achieving the goal reliability includes developing a comprehensive understanding of the performance limitations and possible improvements of the Fermilab Booster.



Following in the tradition of \cite{ShiltsevEldred} and \cite{Huang06}, our approach to a comprehensive understanding of the Fermilab Booster begins in empirical analysis of the performance limits and available beam instrumentation. Next, comes modeling of the relevant physics for emittance growth and beam loss mechanisms, drawing upon international experience with intensity rings with more mature physics models. In this research, we advance the state of knowledge of the Fermilab Booster by sharply highlighting the modeling gaps and identifying appropriate methods for follow-up research.

Table~\ref{tab:BoosterParam} provides an overview of Booster parameters.

\begin{table}[htbp]
\begin{center}
{\begin{tabular}{|l|c|l|}
\hline
Parameter & & Comments \\
\hline 
Circumference, $C$ & 474.20 m & \\
Injection Energy (kinetic), $E_i$	& 400 MeV	& $\beta_p$=0.701, $\gamma_p$=1.426 \\
Extraction Energy (kinetic), $E_f$	& 8 GeV	& $\beta_p$=0.994, $\gamma_p$=9.526 \\
Cycle time, $T_0=1/f_0$ & 	1/15 s & 20,000 revolutions \\
Harmonic number, $h$ &	84	& \\
RF frequency, $f_{RF}$& 37.77-52.81 MHz & inj.-extr.	\\
Max RF voltage, $V_{RF}$ & 1.1 MV	& \\
Momentum compaction, $\alpha_{c}$ & 0.034803 & $\gamma_{tr}$=5.478, at $t$=17ms \\
No. of cells, magnets	& 24, 96 & $FOFDOOD$, 96$^o$/cell \\
Nominal intensity, $N_p$	& 4.5$\cdot 10^{12}$ & $N_b$=81 bunches \\
Rms norm.emitt., $\varepsilon_{x,y}$ & 2.0 $\pi \,\mu$m	& 12 $\pi \, \mu$m for 95 \% extracted \\
Max Betatron functions, $\beta_{x,y}$ &	33.7/20.5 m & ~ \\
Max Dispersion function, $D_x$ &	3.2	m & ~  \\
\hline
\end{tabular}}
\caption{Main operational parameters of the Fermilab Booster. $\beta_p$ and $\gamma_p$ are relativistic Lorentz factors of protons.}
\label{tab:BoosterParam}
\end{center}
\end{table}

\begin{figure}[htbp]
\centering
\includegraphics[width=.6\textwidth]{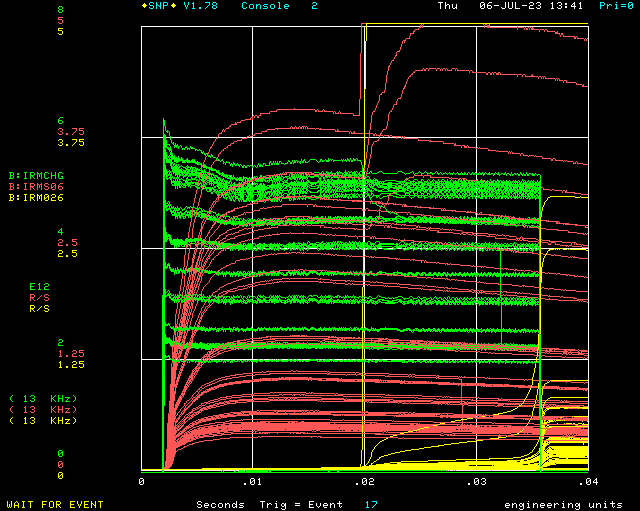}
\caption{In green, the total intensity in the ring measured by the toroid. In red and yellow, loss monitors for locations near to collimation and a vertical aperture restriction near extraction.}
\label{fig:Instability_Elog}
\end{figure}

Recent Booster machine development efforts have been focused on identifying barriers between increasing intensity from its nominal value (4.5e12) to its PIP-II era value (6.5e12). Figure~\ref{fig:Instability_Elog} shows the losses as injected intensity is increased. In the early part of the Booster cycle after RF capture, space-charge forces dominate and the losses increase nonlinearly. Above an intensity threshold of $\sim$5.5e12 and near transition (19.26ms), a fast beam instability takes place. The early space-charge losses will be discussed in Section~\ref{sec:dc} and Section~\ref{sec:2qy}, while the transition instability is investigated in Section~\ref{sec:headtail}. An overview of the Booster instrumentation in provided in \cite{ShiltsevEldred}.

\section{2023 International Fermilab Booster Studies}
\label{sec:studies}

The International Fermilab Booster Studies began in June 2019, after a memorandum of understanding (MOU) between Fermilab and CERN for mutual visits of scientific personnel. The 2019 studies included six topics over three weeks. Active study time included 36 hours of dedicated beam study time and 52 hours of parasitic (simultaneous with HEP operations). In addition to six scientific visitors from CERN, there were also two scientific visitors from US-based research contractor Radiasoft LLC and one scientific visitor from GSI Helmholtz Centre for Heavy Ion Research. An overview of the main results of the studies was presented at a hybrid workshop the following August~\cite{EldredCapstone}. The research output of the 2019 beam studies include a comprehensive study on emittance and loss scaling with beam intensity~\cite{ShiltsevEldred}, a new calibration of Booster ionization profile monitors (IPMs)~\cite{ShiltsevIPM}, an analysis of the susceptibility of Booster gradient magnet circuits to power supply ripple effects~\cite{SchmidtRipple}, a localization of Booster impedance effects~\cite{Biancacci20}, and several areas of continued study that are presented in this manuscript. 

Due to the COVID-19 pandemic, in-person international beam studies did not take place again until the 2023 International Fermilab Booster Studies. Originally the beam study program was planned to take place in five consecutive days June~26th to June~30th 2023 coinciding with the schedule of three visiting scientists from CERN. Due to unexpected operational difficulties, Fermilab was not able to conduct the studies that week. However, dedicated beam study time of the same total duration was able to take place between July 5-11th, 2023 with participation from CERN. Table~\ref{tab:Schedule} shows the duration and timing of each beam study session as it ultimately occurred, for the three study topics.

\begin{table}[htbp]
\begin{center}
\begin{tabular}{|c c c|} 
 \hline
 Date & Time & Type \\
 \hline\hline
 April 26, 2023 & 13:15-16:30 & \textcolor{ForestGreen}{400~MeV Mode Space-charge} \\
 \hline
 June 7, 2023 &  8:30-12:30 & \textcolor{ForestGreen}{400~MeV Mode Space-charge} \\ 
 \hline
 June 7, 2023 & 15:50-17:20; 20:00-22:00 & \textcolor{Red}{2$Q_{y}$ Resonance} \\ 
 \hline
 June 8, 2023 & 8:00-10:00 & \textcolor{ForestGreen}{400~MeV Mode Space-charge} \\ 
 \hline
 July 5, 2023 & 8:00-12:10 & \textcolor{ForestGreen}{400~MeV Mode Space-charge} \\ 
 \hline
 July 6, 2023 & 8:00-11:40 & \textcolor{ForestGreen}{400~MeV Mode Space-charge} \\ 
 \hline
 July 6, 2023 & 12:30-16:45 & \textcolor{Blue}{Transition Instability} \\ 
 \hline
 July 7, 2023 & 8:00-15:00 & \textcolor{Red}{2$Q_{y}$ Resonance} \\ 
 \hline
 July 10, 2023 &  2:45-16:20 & \textcolor{ForestGreen}{400~MeV Mode Space-charge} \\ 
 \hline
 July 11, 2023 & 8:00-14:00 & \textcolor{Red}{2$Q_{y}$ Resonance} \\ 
 \hline
\end{tabular}
\caption{The dates and time each beam study presented in this work. Each of the three study topics is shown in a unique color.}
\label{tab:Schedule}
\end{center}
\end{table}

For the first study topic (Section~\ref{sec:dc}) we placed the Booster into 400-MeV mode and studyied the space-charge effects with loss monitors, IPMs, and simulation code. The second study topic (Section~\ref{sec:2qy}) was an outgrowth of the last several years of departmental Booster studies and focuses on the mitigation of the $2Q_{y}=13$ resonance to improve the performance of high-intensity beam operations. The third study topic (Section~\ref{sec:headtail}) was a continuation of a 2019 study topic and investigates a tail-dominated mode coupling instability with strong space-charge in the vicinity of transition.

\section{Space-charge in 400~MeV Booster}
\label{sec:dc}

\subsection{Beam Loss Mechanism \& 400~MeV Tuning}
\label{sec:dc1}

For the International Booster Studies, the Fermilab Booster was placed in a ``400~MeV'' mode. In 400 MeV mode, the Booster gradient magnets do not ramp at 15~Hz, but rather continuously hold their injection values (nominally 0.078T for F-magnets and 0.066T for D-magnets). The RF cavities use paraphasing for adiabatic capture as usual, after which the phase, frequency, and voltage is not changed because the beam is not accelerated. Although the beam is stored for the entire 66~ms cycle, the RF is limited in duty factor and deactivated at 33~ms (as it would for a ramping cycle). The 48 multipole corrector packages are held constant at their injection values, and are not ramped. The size of the injected beam and the pulse duration of the extraction kickers are not appropriate for extraction to the 8~GeV dump, and consequently the injected beam is simply lost in the ring at 400~MeV.

The quadrupole and sextupole components of the Booster magnets appear to be impacted by the change from the ramping mode to the 400~MeV mode. At the start of a ramping cycle, the betatron tunes are $Q_{x}=6.75,Q_{y}=6.85$ whereas in the 400~MeV cycle, the betatron tunes are $Q_{x}=6.67,Q_{y}=6.82$ . Similarly, at the start of a ramping cycle, the chromaticities are $Q_{x}^{\prime}=-17,Q_{y}^{\prime}=-5$ whereas in the 400~MeV cycle, the chromaticities are $Q_{x}^{\prime}=-15,Q_{y}^{\prime}=0$ . To stabilize the beam, it was necessary to strengthen the horizontal chromaticity, so $Q_{x}^{\prime}=-17,Q_{y}^{\prime}=0$ was used for the 400~MeV studies.

\begin{figure}
  \centering
  \subcaptionbox{In green, the total intensity in the ring measured by the toroid. In red, yellow, and blue, loss monitors for locations near to collimation, horizontal aperture restriction, and sensitive RF cavities respectively. \label{fig:lifetime_A}}
    {\includegraphics[width=.4\linewidth]{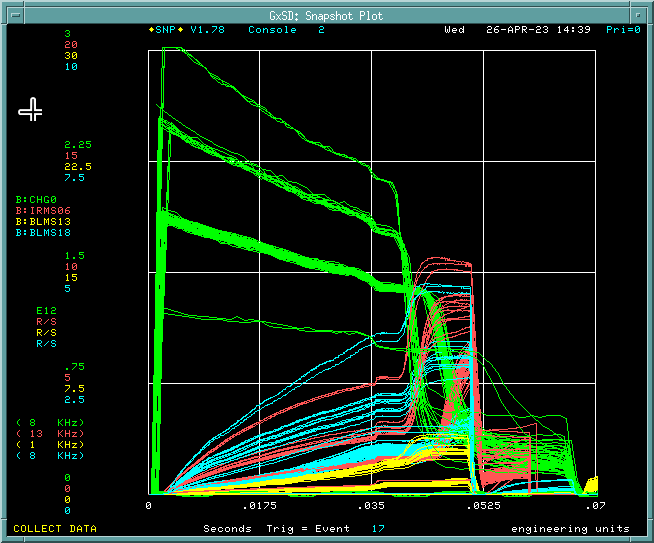}}\qquad
  \subcaptionbox{Losses per unit time (e11/ms) as a function of beam intensity, indicating a roughly quadratic dependence. The loss mechanism is not impacted by time within cycle (before 35ms) indicated by the color scale. \label{fig:lifetime_B}}
    {\includegraphics[width=.4\linewidth]{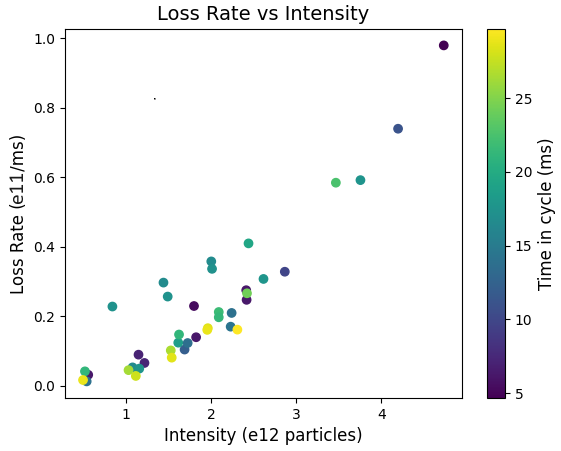}}
    \caption{Time dependence and intensity dependence of loss mechanism in 400~MeV operation. \label{fig:lifetime}} 
\end{figure}

In Figure~\ref{fig:lifetime_A} an Accelerator Network (Acnet) ''Snapshot'' plot is shown depicting measured charges and losses for several 400~MeV Booster cycles of four different initial intensities. A steady intensity-dependent loss mechanism is observed until 35ms into the cycle, when RF bunching is turned off, at which time a fast beam instability takes over. 

The nature of the loss mechanism is still being investigated and this work is the first time its features have been characterized in 400~MeV Booster cycles. In Figure~\ref{fig:lifetime_B} the loss rate measured by the toroid for several cycles. The loss rate grows nonlinearly with beam intensity indicating a charge-dependence to the loss mechanism (i.e. rather than a constant loss per particle characteristic of an exponential decay). Another notable feature is the absence of intensity thresholds or sigmoid loss patterns; instead the loss mechanisms has indications of a slow, diffusive or noise-driven process.

\begin{figure}[htbp]
\centering
\includegraphics[width=.6\textwidth]{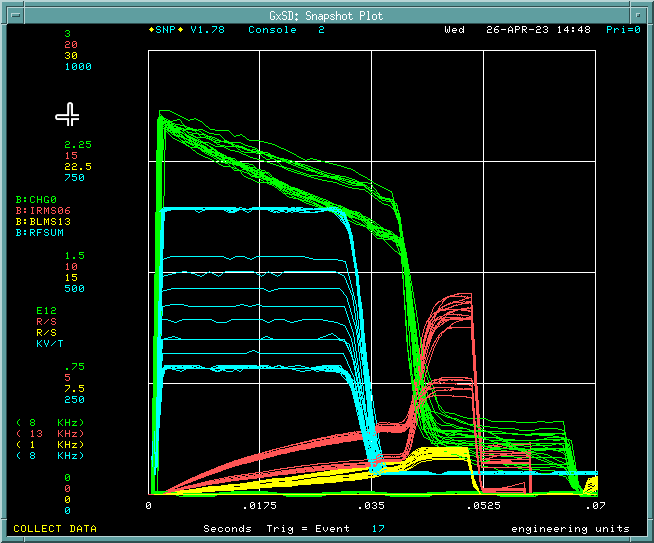}
\caption{In green, toroid measurement of charge. In red and yellow, loss monitors for collimation and horizontal aperture locations. In blue, the total RF voltage across all RF cavities.}
\label{fig:lifetimeRF}
\end{figure}

At first glance, the loss mechanism is not strongly impacted by small changes in beam tune, although a comprehensive investigation of tune-dependence for low-intensity beams is presented in the next subsection. The loss rate was found to be sensitive to the magnitude of chromaticity and RF bunching. Figure~\ref{fig:lifetimeRF} shows a Snapshot plot for several 400~MeV Booster cycles at two different total RF voltages. A $\sim$50\% reduction in the RF voltage leads to a $\sim$40\% in the loss-rate. Reducing the RF voltage has the effect of reducing the space-charge tune-spread, reducing the chromatic tune-spread, and reducing the synchrotron tune - in this preliminary study we were not able to differentiate which effect moderates the loss mechanism.

Presuming the loss-rate per particle per unit time is a function of tune-shift and chromaticity, the empirically observed 400~MeV loss rate cannot be fully reconciled with ramping loss rate. From Figure~\ref{fig:lifetime_B} we have
\begin{align} \label{eq:loss_dc}
\frac{1}{N} \frac{d N}{dt} \approx 0.02~ms^{-1} \left( \frac{N}{5\text{e12}} \right) \left \vert \frac{Q_{x}^{\prime}+Q_{y}^{\prime}}{17} \right \vert^{\alpha}
\end{align}
for some unknown $\alpha$, whereas in \cite{ShiltsevEldred}, a study of the loss rate in a ramping Booster cycle was found to be:
\begin{align} \label{eq:loss_ramp}
\int \frac{1}{N} \frac{d N}{dt} dt \approx 0.013 + 0.027 \left( \frac{N}{5\text{e12}} \right)^{3} \left \vert \frac{Q_{x}^{\prime}+Q_{y}^{\prime}}{17} \right \vert ^{1.9}
\end{align}
Although the functional forms differ, we can calculate the timescale where losses are comparable to be $\sim$2~ms for a beam intensities between 2e12 and 5e12. This is broadly consistent with the observation that space-charge losses in the ramping cycles are dominated by the lowest energies where the space-charge tune-shift is large. However, in Section~\ref{sec:dc4} we show that the beam emittance changes dramatically while the 400~MeV loss rate remains steady, suggesting the losses are not a simple function of the space-charge tune-shift. The Booster has significant dispersion in its RF cavities, and subsequent work will investigate whether syncho-betatron resonances play a role in the loss mechanism that is subsequently moderated by acceleration.



\subsection{Low Brightness Tunescan at 400~MeV}
\label{sec:dc2}

Tunescans (measuring losses as a function of betatron tune) are the gold standard measurement tool for the identification of dangerous resonances in betatron motion. Tunescans for ramping operations of the Fermilab Booster only show indications of first and second-order resonances. Recent work at the CERN PS~\cite{Asvesta20} has demonstrated that the identification of higher-order lattice resonances are enhanced by the use of ``low-brightness'' beams which feature more particles at larger betatron amplitudes. In this study, the technique is applied for the first time in the Fermilab Booster and applied to stable 400-MeV cycles.

In the Fermilab Booster, the tunescan measurement is calculated from the ratio of the beam intensity after 3ms to the injected beam intensity, and plots that ratio as a function of the horizontal and vertical tune. The tunes are changed according to the normal procedure of tune control in the Booster, in two families of 24 changed symmetrically across all Booster cells. The plots that follow change the tunes in 0.025 unit increments, with $Q_{x}$ from 6.52 to 6.87 and $Q_{y}$ from 6.52 to 6.92. The tune is not measured independently for each point, but rather calculated from the quadrupole changes deviating from a single point with measured tunes ($Q_{x}$=6.67, $Q_{y}$=6.82). Only two turns were injected, corresponding to about 0.55e12 protons, to minimize space-charge forces and activation.

\begin{figure}[htbp]
\centering
\includegraphics[width=.6\textwidth]{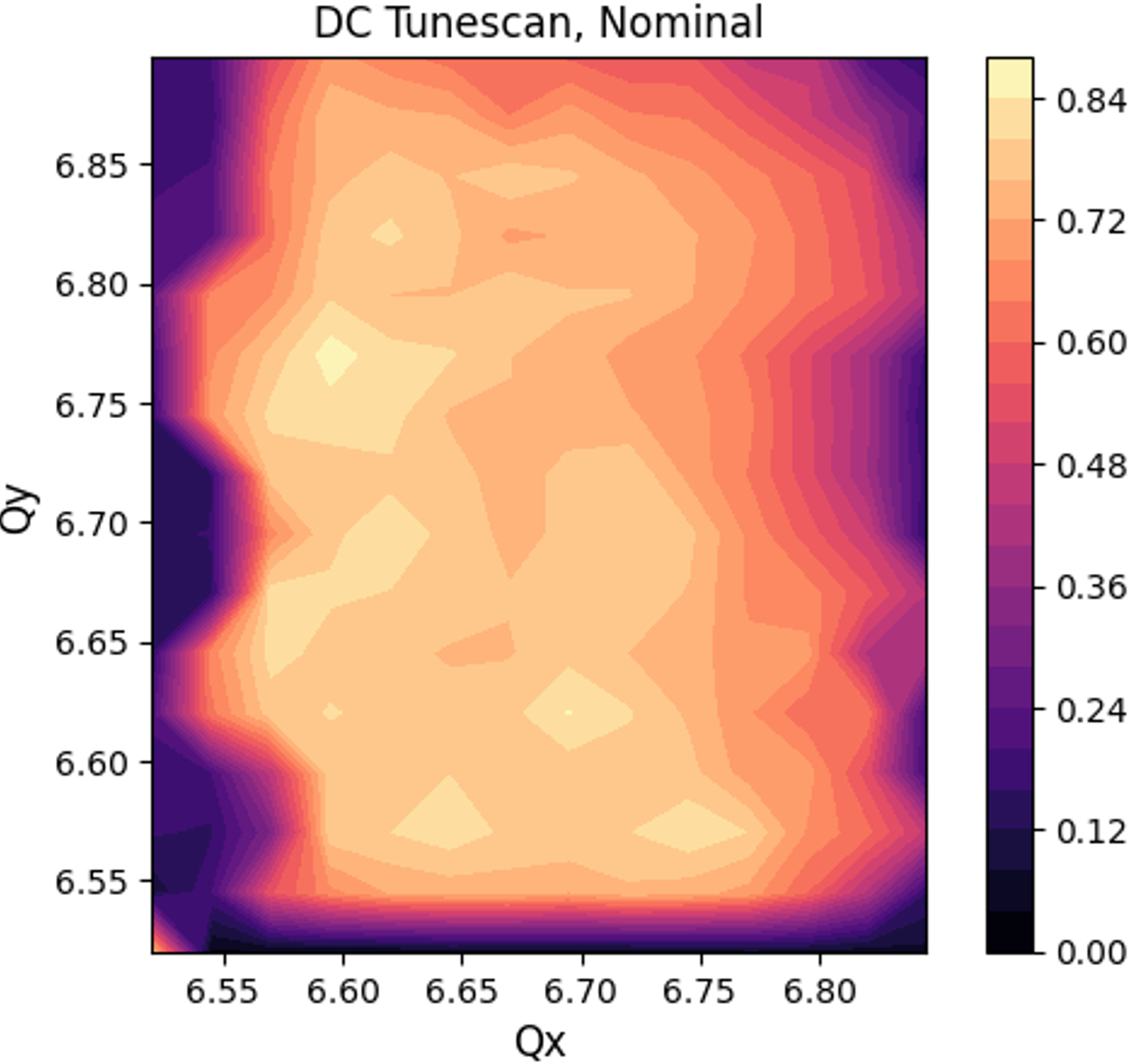}
\caption{Fraction of surviving beam as a function of tune, with nominal conditions.}
\label{fig:TS}
\end{figure}

Figure~\ref{fig:TS} shows the tunescan under nominal operating conditions for the 400~MeV cycle. The half-integer (${2Q_{y}=13}$ and ${2Q_{x}=13}$) and integer stopbands (${Q_{y}=7}$ and ${Q_{x}=7}$) lead to catastrophic beam loss, while no other resonance lines are clearly observed. A faint ${3Q_{x} = 20}$ resonance line may be present, and there may be additional losses in the proximity of half-integer and integer resonances where other resonances cross.

The Booster correctors were left at the their operational values, which are based on empirically tuning for the best loss profiles, rather than direct minimization of any specific resonances. However, this blind optimization procedure has the effect of roughly minimizing the three nearby sextupole resonances: ${3Q_{x}=20}$, ${Q_{x}+2Q_{y}=20}$ and ${Q_{x}-2Q_{y}=-7}$. For some tunescan measurements one of the 48 sextupole correctors, ``S19'' (for its location at the short straight of Booster cell 19) was increased by 4 amps to deliberately excited a sextupole perturbation. The S19 sextupole corrector has a horizontal beta function $\beta_{x} \approx 35~m$, a vertical beta function $\beta_{y} \approx 5~m$, and at 4 amps has an integrated sextupole kick $K_{2}L = 0.115 m^{-2}$.

Two mismatch procedures were developed to create low-brightness beams in the Fermilab Booster in order to conduct more sensitive tunescans. In the horizontal mismatch, a 9.65~mm horizontal 4-bump was applied during injection, painting the beam with a 4.2 sigma horizontal offset. In the vertical mismatch, a 10.06~mm vertical 4-bump was applied during injection, painting the beam with a 5.4 sigma vertical offset. In both cases, the mismatch conditions resulted in $\sim$15-30\% of the beam immediately lost due to scraping effects.

\begin{figure}
  \centering
  \subcaptionbox{Fraction of surviving beam as a function of tune, with sextupole perturbation. ${3Q_{x}=20}$ line indicated. \label{fig:TS_S}}
    {\includegraphics[width=.4\linewidth]{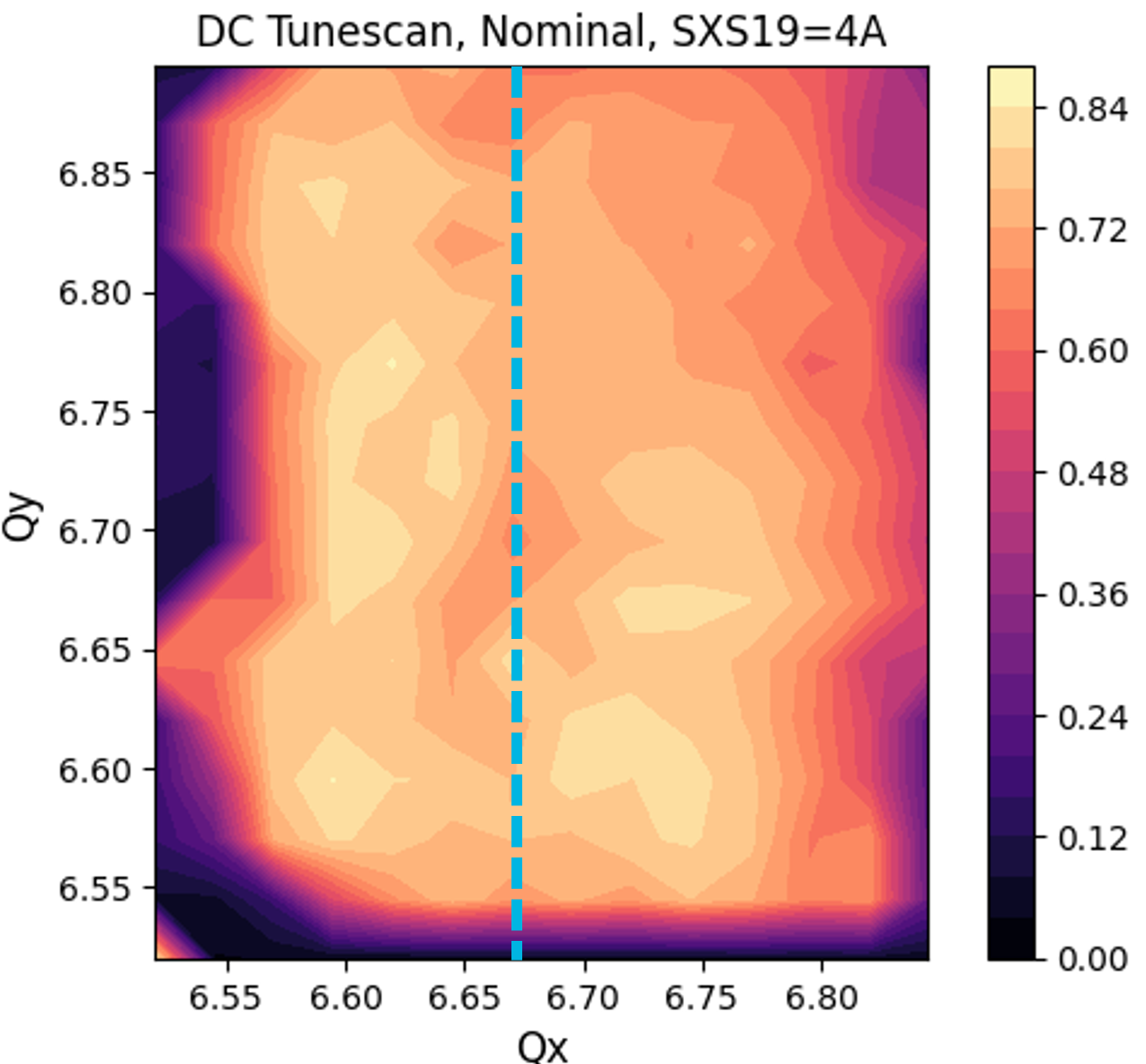}}\qquad
  \subcaptionbox{Fraction of surviving beam as a function of tune, with horizontal mismatch and sextupole perturbation. $3Q_{x}=20$ line indicated. \label{fig:TS_HS}}
    {\includegraphics[width=.4\linewidth]{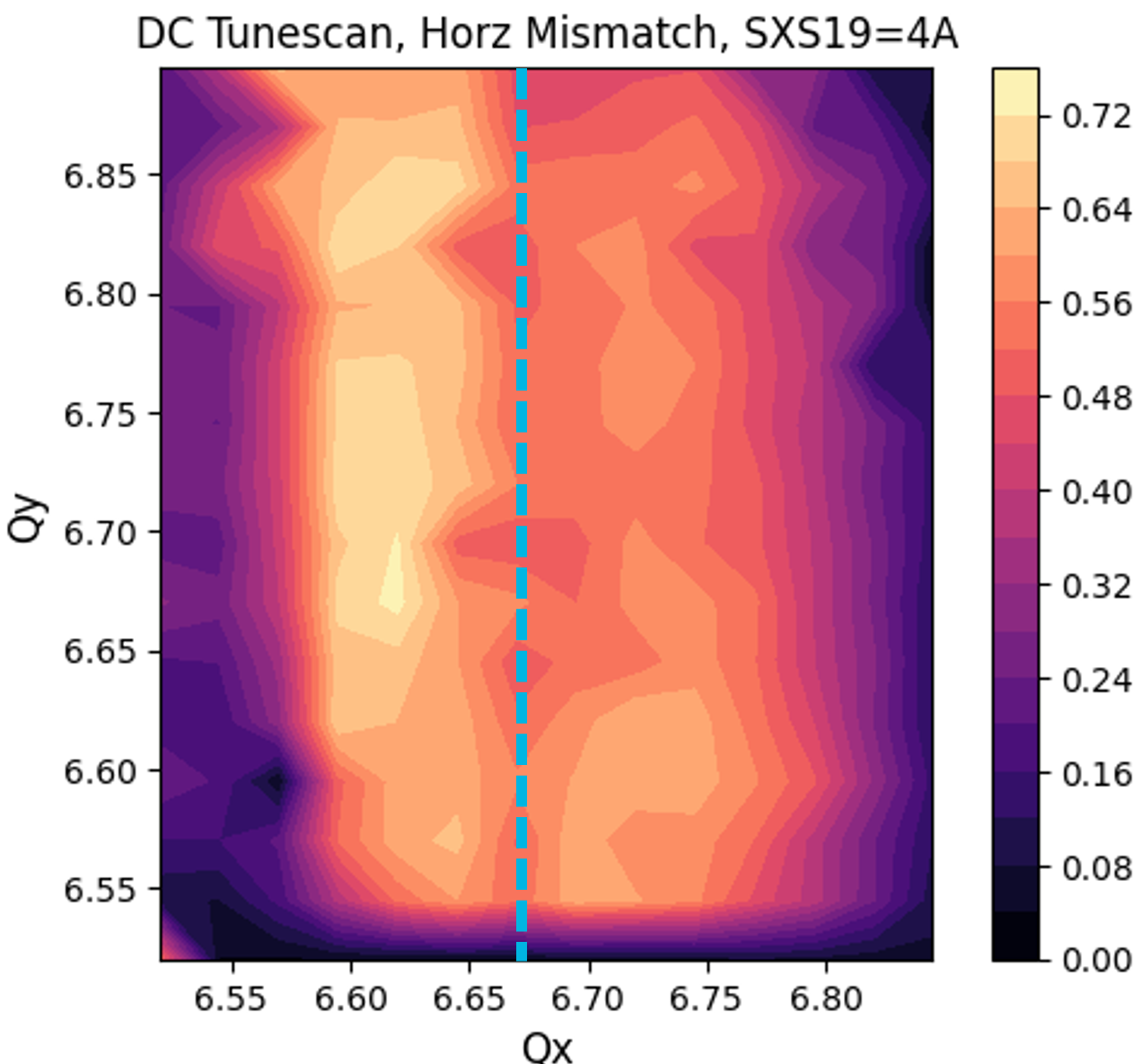}}
    \caption{Impact of 4A sextupole perturbation on 400~MeV tunescans \label{fig:TS_H_HS}} 
\end{figure}

Figure~\ref{fig:TS_S} shows the tunescan with 4 amp sextupole perturbation at the S19 corrector. Enhanced losses are clearly observed along the ${3Q_{x} = 20}$ resonance line as a result of the sextupole perturbation. Figure~\ref{fig:TS_HS} shows the tunescan with the combination of the sextupole perturbation and the horizontal mismatch conditions. In that case, excess beam losses also take place in the region above the ${3Q_{x} = 20}$ resonance line.

\begin{figure}
  \centering
  \subcaptionbox{Fraction of surviving beam as a function of tune, with vertical mismatch. ${Q_{x}-Q_{y}=0}$ line indicated. \label{fig:TS_V}}
    {\includegraphics[width=.4\linewidth]{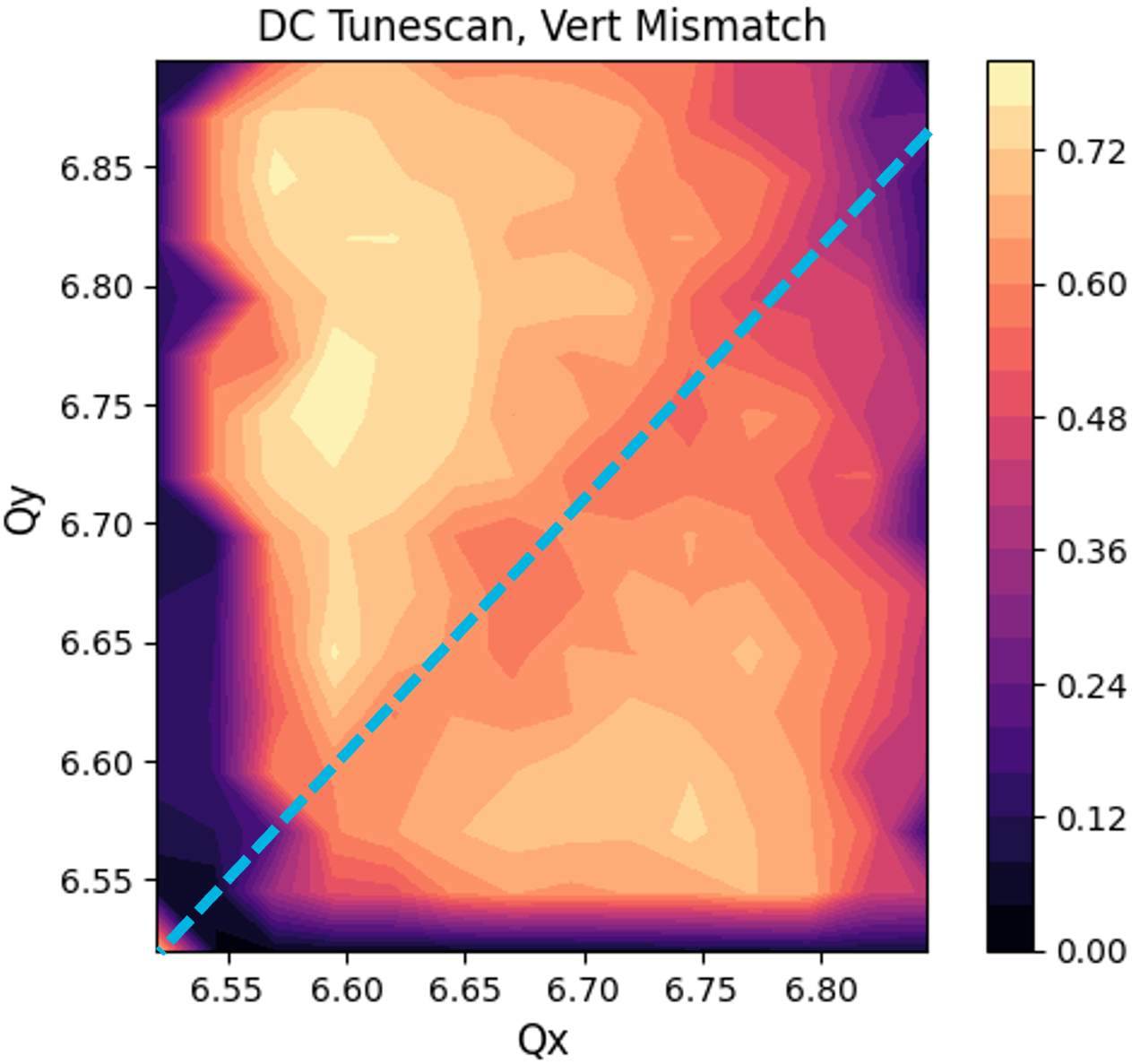}}\qquad
  \subcaptionbox{Fraction of surviving beam as a function of tune, with vertical mismatch and sextupole perturbation. ${3Q_{x}=20}$ and ${Q_{x}-Q_{y}=0}$ lines indicated. \label{fig:TS_VS}}
    {\includegraphics[width=.4\linewidth]{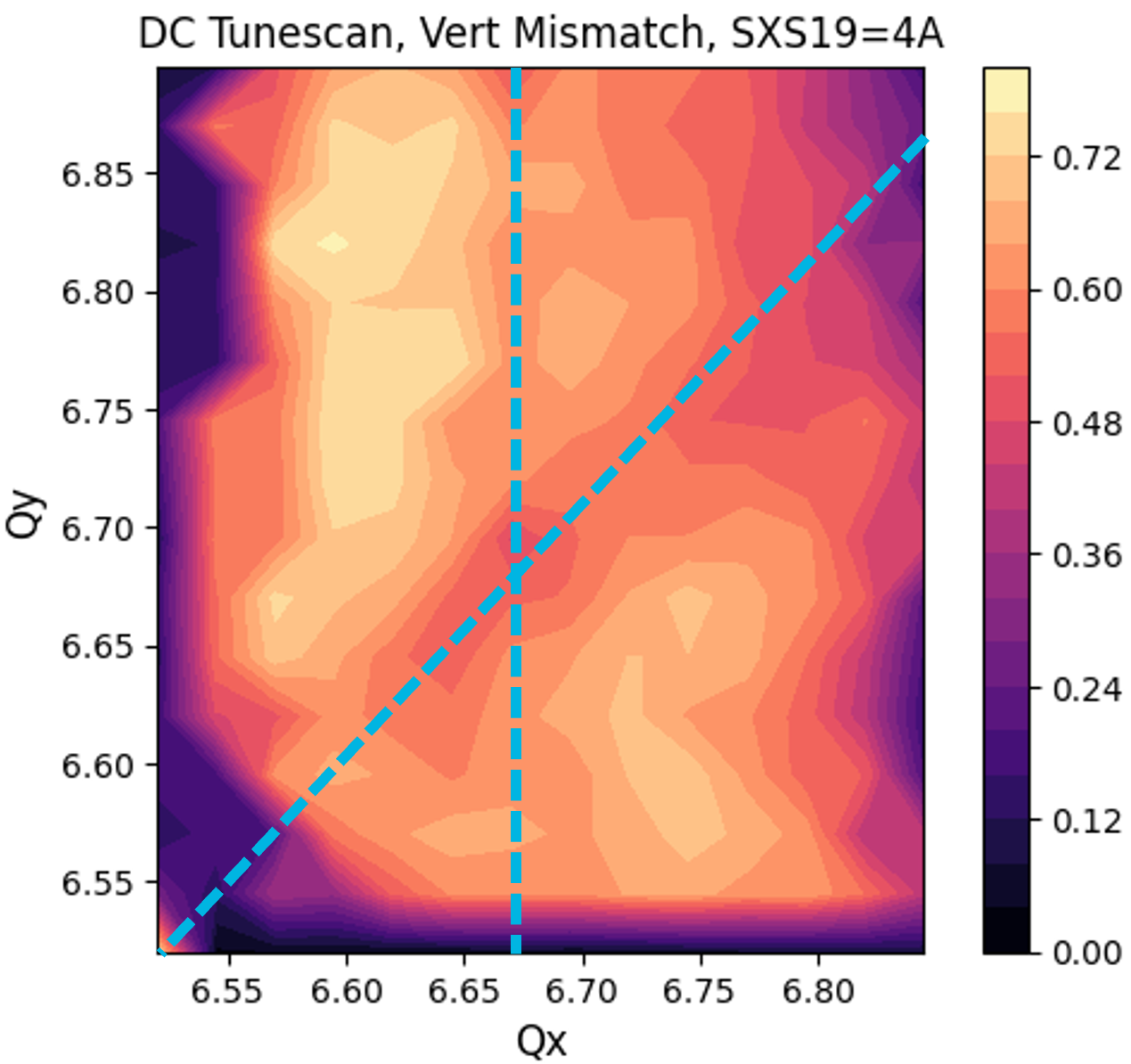}}
    \caption{Impact of vertical mismatch on 400~MeV tunescans \label{fig:TS_V_VS}} 
\end{figure}

Figure~\ref{fig:TS_V} shows the tunescan under the vertical mismatch condition, and the $Q_{x}-Q_{y} = 0$ coupling line is clearly seen. Figure~\ref{fig:TS_VS} shows the tunescan under the combination of the sextupole perturbation and the vertical mismatch conditions. In that case, both the $Q_{x}-Q_{y} = 0$ resonance line and the $3Q_{x} = 20$ resonance line can be clearly seen.

The presence of linear coupling line is not altogether surprising. Although the Booster has skew quadrupole correctors that can eliminate global coupling, it is deliberately operated with weak linear coupling (closest tune difference of $\sim$ 0.05), to help suppress coherent dipole instabilities in the early part of the cycle~\cite{Alexahin12}. As a result of that linear coupling, when the beam is vertical mismatched, tunes on the coupling resonance may deteriorate the horizontal beam quality and therefore drive additional losses.

The low-brightness tunescans of the 400~MeV Booster were consistent with previous tunescans of the ramping Booster cycle, which showed that half-integer and integer resonances are the dominant resonances for the Fermilab Booster. The impact of the sextupole perturbation on the ${3Q_{x} = 20}$ resonances underscores the soundness of the tunescan methods for detecting higher order resonances, and confirms that the available sextupole correctors are generally effective at suppressing third-order resonances for the Booster. The low-brightness tunescan method is not setup to investigate space-charge induced resonances, which likely play a significant role in Booster performance apart from the dipole and quadrupole resonances.

\subsection{MADX-SC Simulation of 400~MeV Beam}
\label{sec:dc3}

Recently, an adaptive space-charge model ``MADX-SC''~\cite{Schmidt:2644660} has been incorporated into the official CERN version of the MAD-X particle tracking code~\cite{riccardo_de_maria_2023_7900976}. During particle tracking each revolution, the MADX-SC code calculates the beam intensity and beam size in order to parameterize Gaussian space-charge kick elements distributed around the ring. Coinciding  with the LHC Injector Upgrade~\cite{Damerau:1976692} (LIU) brightness improvement campaign, the MADX-SC code has been successfully benchmarked against machine performance measurements in the CERN PS~\cite{PhysRevAccelBeams.20.081006}. The core of the benchmarked effort is to predict the evolution of the transverse profiles as a function of beam intensity and other machines parameters.

The first step of applying the simulation to the Fermilab Booster is to identify the correct machine parameters. In addition to the parameters given in Table~\ref{tab:BoosterParam} above, Table~\ref{tab:dcParam} and Table~\ref{tab:SC_Tunes} give parameters used for the 400~MeV beam studies and simulations. The initial normalized RMS emittance used for simulation is based on wirescanner measurements in the linac before injection. The total RF voltage was specifically reduced to manage the losses described in Section~\ref{sec:dc1} and the corresponding longitudinal parameters are given. Stripline measurements and longitudinal calculations indicate that some of the beam may not be captured at this voltage, but the bunching structure predominates. For simulation, the longitudinal distribution of the bunch was populated with phase truncated to $\pm 2 \sigma$ to avoid populating particles outside of the RF bucket. In simulation, machine apertures are not used.

\begin{table}[htbp]
\begin{center}
{\begin{tabular}{|l |l|}
\hline
Initial norm. RMS emittance & 1.6~$\pi$~mm~mrad \\
Total RF Voltage & 250~kV \\
RMS Momentum Spread & 1.52e-3 \\
Bunch Length & 5.18~ns \\
Bare Tunes $Q_{x}, Q_{y}$ & 6.67, ~6.82 \\
Chromaticity $Q_{x}^{\prime}, Q_{y}^{\prime}$ & -17, ~0 \\
\hline
\end{tabular}}
\caption{Parameters used for 400~MeV cycle beam studies.}
\label{tab:dcParam}
\end{center}
\end{table}

\begin{table}[htbp]
\begin{center}
{\begin{tabular}{|l | l | l | l|}
\hline
~ & ~ & 1.6~$\pi$~mm~mrad & 1.0~$\pi$~mm~mrad \\
Case & Intensity & ($Q_{x},Q_{y}$) & ($Q_{x},Q_{y}$) \\
\hline
\hphantom{1}6 turns & 1.7e12 & (6.56, 6.71) & (6.54, 6.68) \\
12 turns & 3.4e12 & (6.44, 6.59) & (6.34, 6.52) \\
18 turns & 5.1e12 & (6.32, 6.47) & (6.19, 6.39) \\
\hline
\end{tabular}}
\caption{Value of space-charge depressed tunes for varying intensities, for the case of initial normalized rms emittance with nominal value (1.6~$\pi$~mm~mrad) and with reduced value (1.0~$\pi$~mm~mrad).}
\label{tab:SC_Tunes}
\end{center}
\end{table}

\begin{figure}
  \centering
    {\includegraphics[width=.45\linewidth]{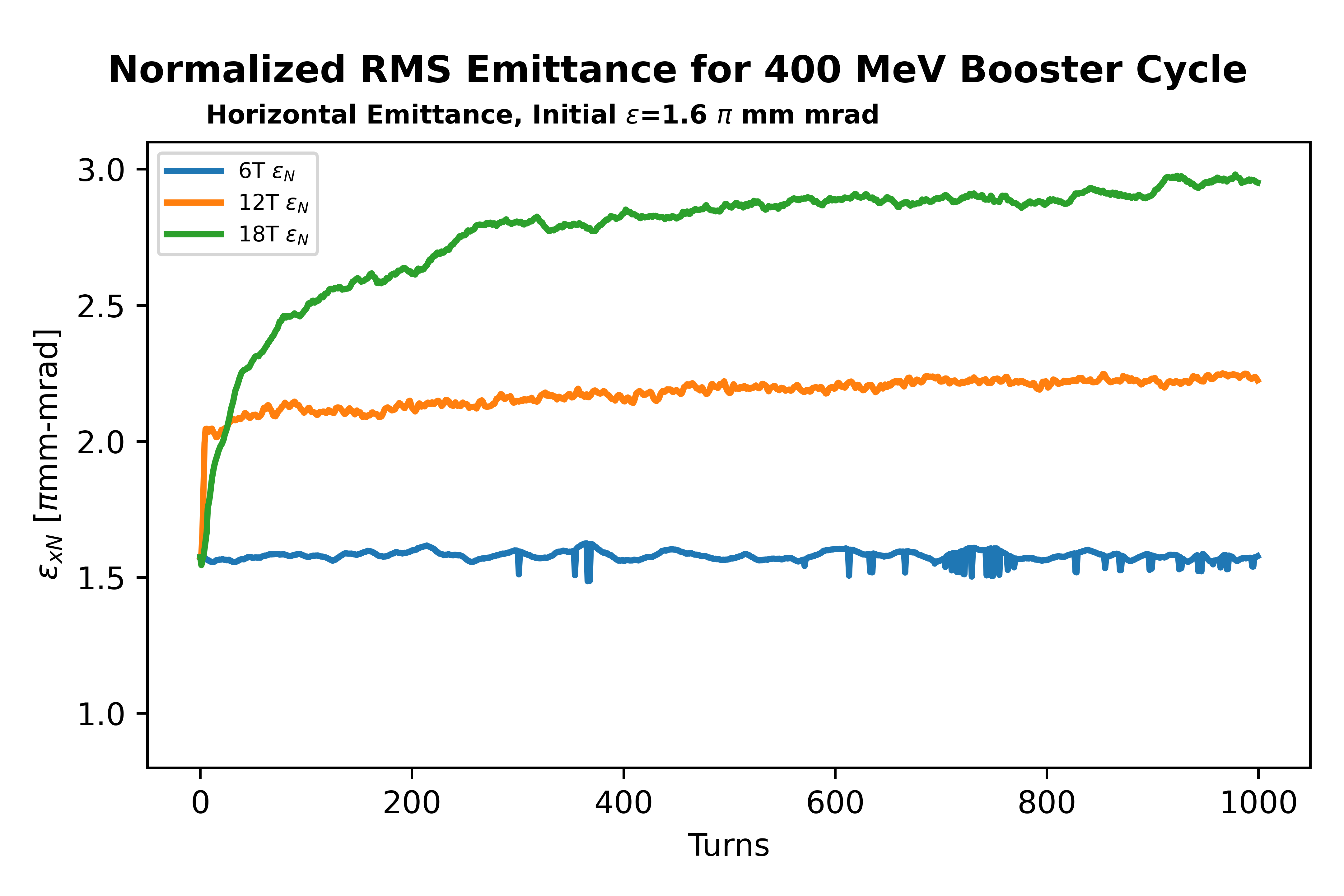}}\qquad
    {\includegraphics[width=.45\linewidth]{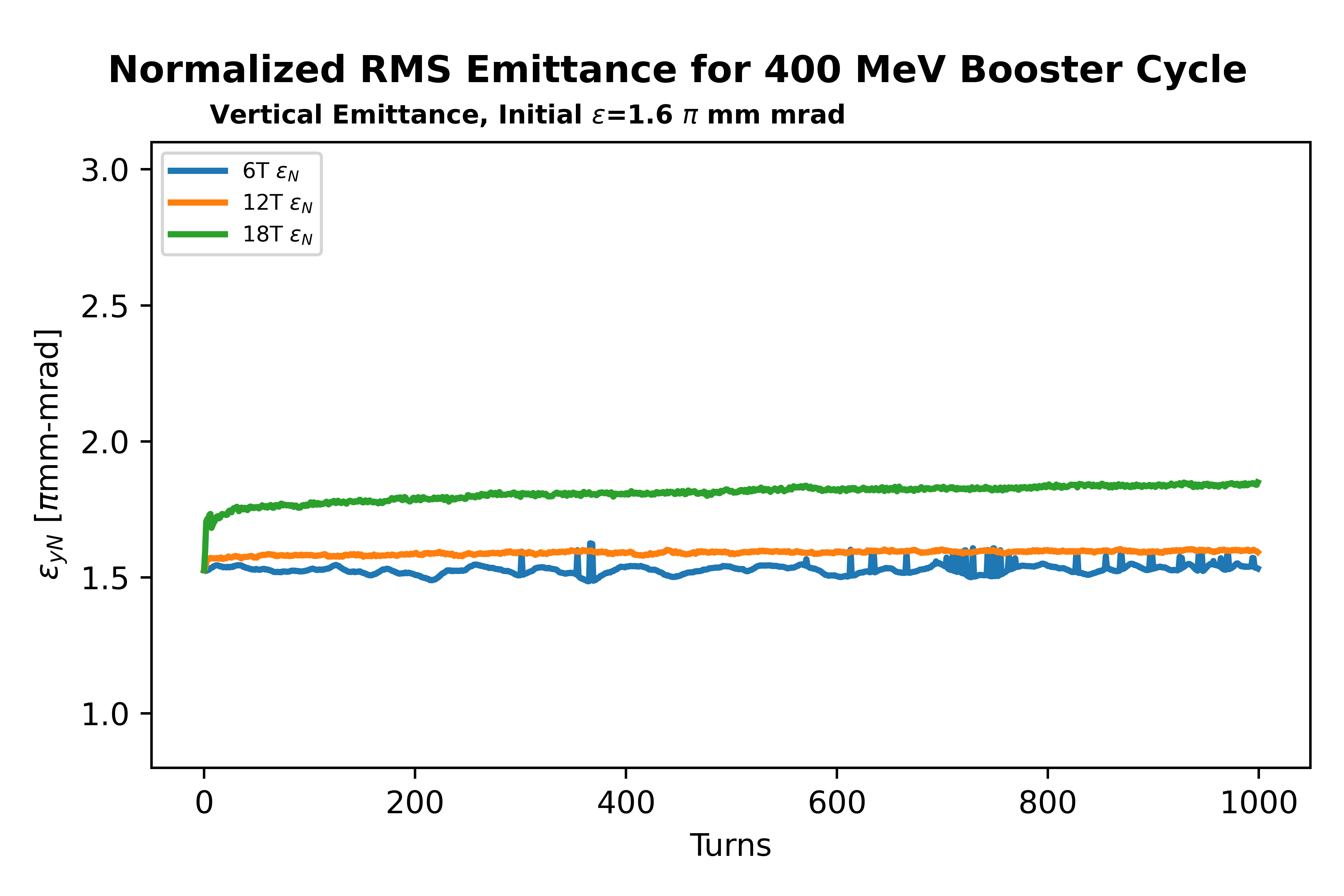}}
    \caption{Evolution of horizontal (left) and vertical (right) normalized rms emittance simulated in MAX-SC code, for nominal initial emittance of 1.6~$\pi$~mm~mrad and varying intensities. \label{fig:MADX}}
\end{figure}

\begin{figure}
  \centering
    \includegraphics[width=.6\linewidth]{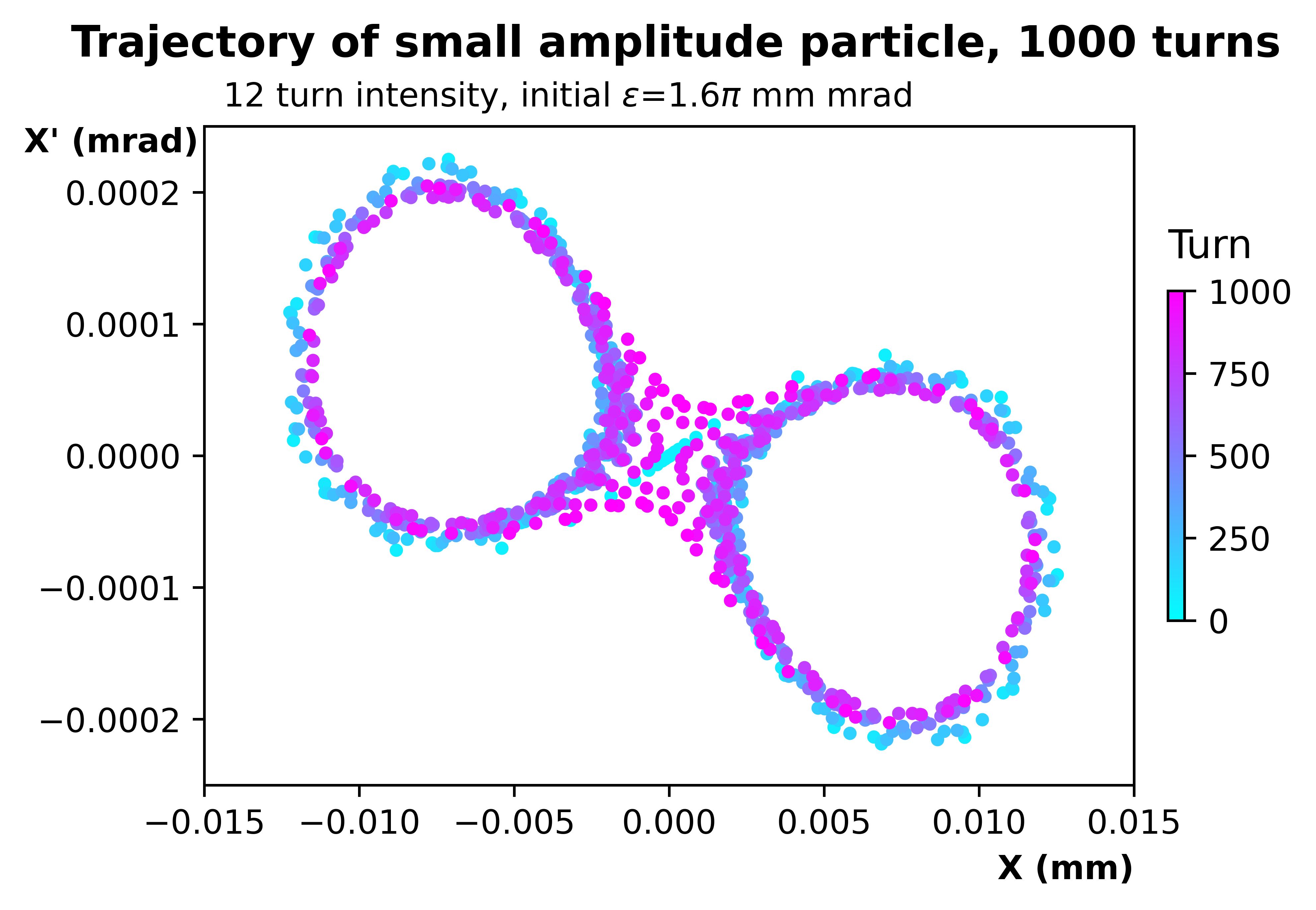}
    \caption{First 1000 revolutions of the horizontal trajectory of a small amplitude particle. 
    \label{fig:trajectory}}
\end{figure}

Figure~\ref{fig:MADX} shows the horizontal and vertical emittance of the MADX-SC simulation computed by SUSSIX code~\cite{Bartolini:1997np} for each intensity case. In the 6 turn case, the horizontal emittance and vertical emittance are relatively constant. In the 12 turn case, the horizontal emittance has a unstable initial value, caused by the fact that the space-charge depressed horizontal tune ($Q_{y}=6.44$) is near the half-integer resonance. Figure~\ref{fig:trajectory} shows the horizontal particle motion near the core of the beam in this 12 turn case, which clearly shows the unstable orbit. In the 18 turn case, the horizontal emittance grows rapidly over 500 turns, and the vertical emittance shows a small unstable shift in emittance (where $Q_{y}=6.47$).

\begin{figure}
  \centering
    {\includegraphics[width=.45\linewidth]{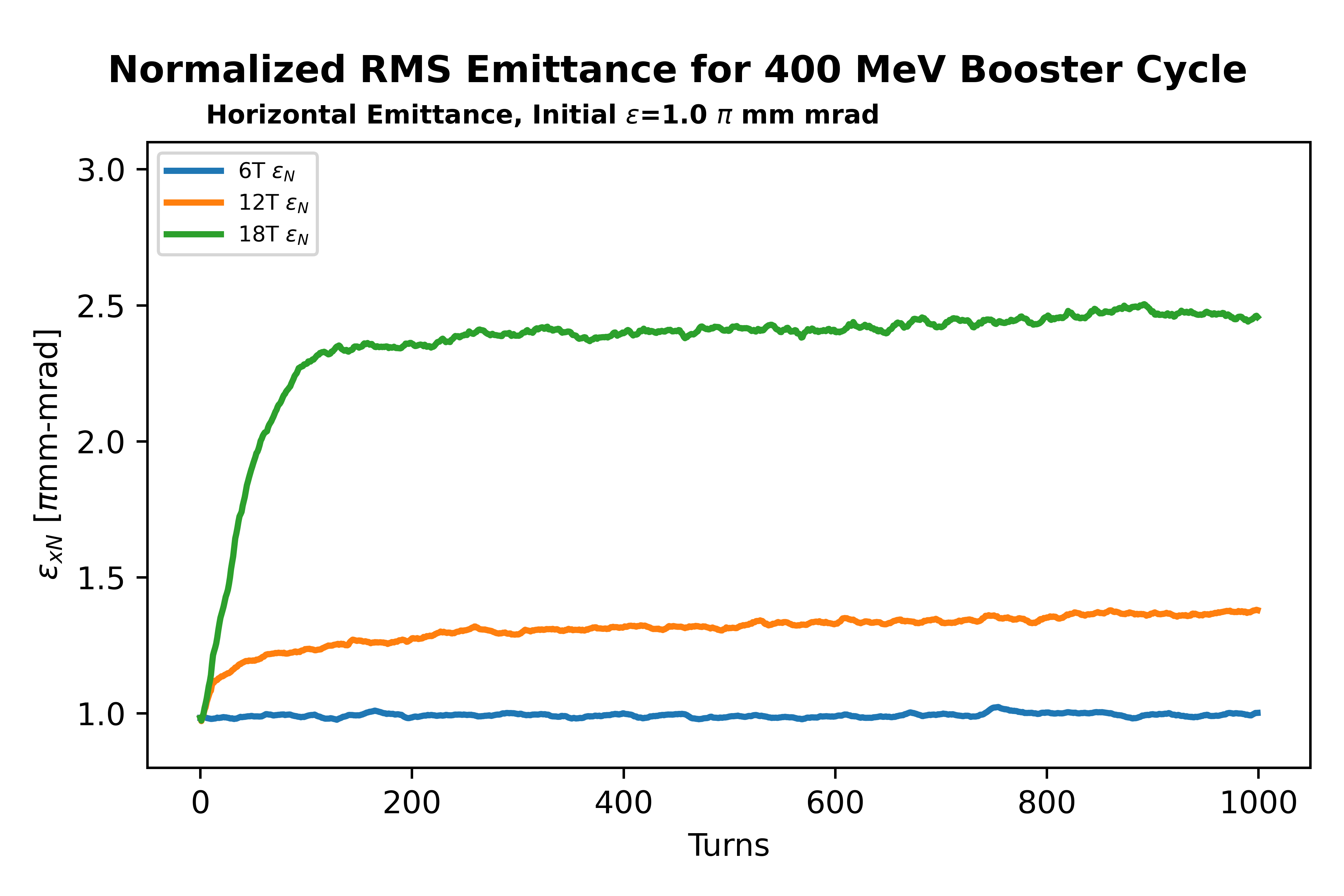}}\qquad
    {\includegraphics[width=.45\linewidth]{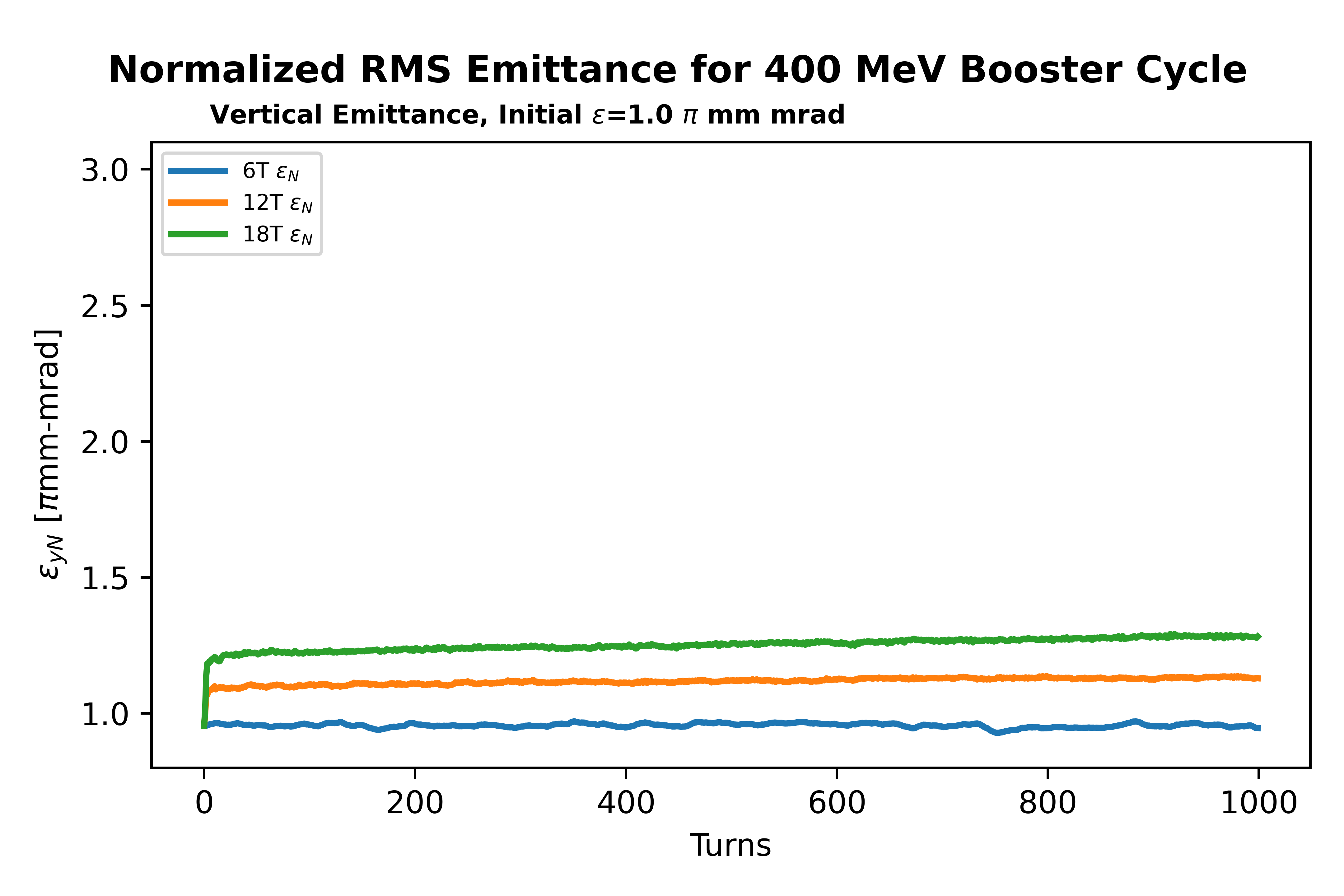}}
    \caption{Evolution of horizontal (left) and vertical (right) normalized rms emittance simulated in MAD-X SC code, for reduced initial emittance of 1.0~$\pi$~mm~mrad and varying intensities. \label{fig:MADX_2}}
\end{figure}

To investigate the behavior of MADX-SC code when the space-charge tune-spread touches or spans the half-integer resonance line, the simulations were also conducted with a reduced initial emittance of 1.0 $\pi$ mm mrad. Figure~\ref{fig:MADX_2} shows the horizontal and vertical emittance of the MADX-SC starting from a reduced initial emittance value. For the horizontal emittances, there are no instances of unstable initialization of emitance. Surprisingly, when the 18 turn horizontal emittance starts at 1.0 $\pi$ mm mrad it reaches an equilibrium at 2.4 $\pi$ mm mrad, but when the value starts at 1.6 $\pi$ mm mrad it instead reaches 3.0 $\pi$ mm mrad. For the vertical emittances, there are unstable initialization in the 12 turn and 18 turn case where the vertical tunes correspond to $Q_{y}=6.45$ and $Q_{y}=6.27$ respectively.

The MADX-SC clearly shows the emittance blow-up mechanism caused by the interaction between space-charge and the $2Q_{y}=13$ resonance (see Section~\ref{sec:2qy} for more discussion). When the space-charge depressed tune is near the half-integer, the MADX-SC code shows the emittance value is immediately unstable as we might expect. 

However, when the space-charge tune-spread spans the half-integer resonance line, the MADX-SC code confounded our expectations. If there is an equilibrium emittance for the space-charge dominated beam, than the reduced initial emittance is more severely mismatched than the nominal initial emittance. Either the two should converge to the same value, or the reduced initial emittance case should suffer a greater deterioration in beam quality.  In \cite{Oeftiger22}, adaptive frozen space-charge codes were found to be consistent with the results of PIC simulations, except when closely approaching resonances. We have begun a collaboration with Lawrence Berkeley National Laboratory to conduct PIC simulations of these non-equilibrium space-charge scenarios using the IMPACT code~\cite{Priv_Comm_Person_MADX-SC-IMPACT}. Additionally, as we will argue in Section~\ref{sec:dc4}, a better understanding of Booster quadrupole errors is likely also needed, following up on previous LOCO analysis presented in \cite{Meghan-McAteer-dissertation}.

\subsection{IPM Measurements of 400~MeV Beam}
\label{sec:dc4}

The Fermilab Booster uses ionization profile monitors (IPMs) to measure the turn-by-turn beam profile. In the case of the Fermilab Booster, ions are collected onto microchannel plates (MCPs) by a 24kV electric field. These measured beam profiles can be distorted by the drift velocity of the ions as well as distortion by the space-charge forces of the proton beams. In \cite{ShiltsevIPM}, using data from the 2019 International Fermilab Booster Studies, a model for ion drift and space-charge effects is developed. With the drift \& space-charge calibration method, the true beam emittance can be more accurately reconstructed from the measured beam profile.

Subsequent investigations have shown that the measured profile sizes are also sensitive to change in the beam orbit and bunch length, but in the 400~MeV studies these parameters are held constant. Typically, horizontal and vertical IPM data is available, however the performance of the horizontal IPM was unreliable at the time of the study.

\begin{figure}[htbp]
\centering
\includegraphics[width=.8\textwidth]{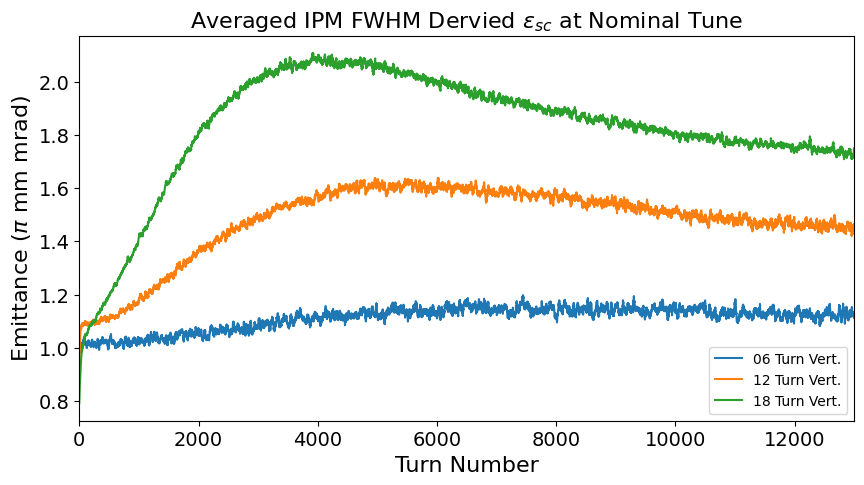}
\caption{RMS normalized beam emittance with drift \& space-charge calibration, for three beam intensities.}
\label{fig:IPM_Nom}
\end{figure}

Figure~\ref{fig:IPM_Nom} shows the calculated vertical beam emittance at the three beam intensities (where 6 turn = 1.7e12, 12 turn = 3.4e12, 18 turns = 5.1e12). The measurements are taken at nominal tune conditions, without sextupole perturbation or mismatch conditions. For the highest two intensity cases, the calculated emittance peaks and begins to decrease as a result of particles lost from the core to the edge of the beam (including losses).

Comparing Figure~\ref{fig:IPM_Nom} to Figure~\ref{fig:MADX} above, we see very significant discrepancies. Both figures show approximately beam emittances on roughly the same scale and find that as the intensity increases the equilibrium emittance increases. In the MADX-SC simulation, the vertical emittance doesn't change much, whereas the IPM emittances show a large change. The horizontal emittance changes very rapidly in the MADX-SC simulation, reaching equilibrium after about 500 turns, whereas the vertical emittance measured by the IPM peaks at 4000 turns and continues to evolve. On the scale of 500 turns, less than 2\% of the beam has been lost, so the discrepancy between simulation and measurement is not due to scraping effects.

Between the IPM data and the unexpected beam loss described in Section~\ref{sec:dc1}, we conclude there is an unmodeled effect driving the emittance growth and particle loss. In \cite{Forte16,Meghan-McAteer-dissertation}, small quad errors in the CERN PSB lattice model were found to generate dramatic effects. But on the other hand, the knowledge of these errors allowed to faithfully predict, with simulations, the losses measured in the CERN PSB. Presently, there are efforts underway in the Fermilab Booster to incorporate misalignment effects and also to improve modeling of the Booster gradient magnet field multipoles.

\begin{figure}[H]
  \centering
    {\includegraphics[width=.8\linewidth]{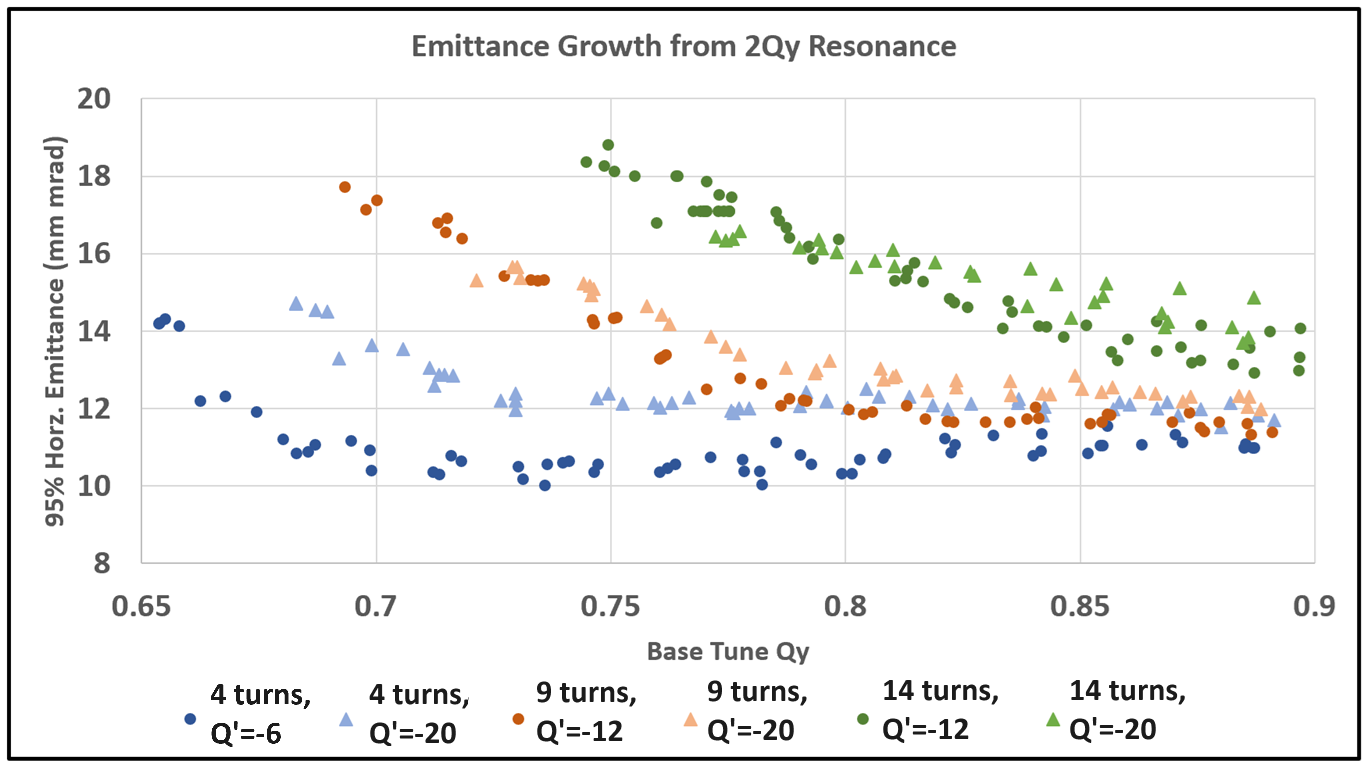}}\qquad
    {\includegraphics[width=.8\linewidth]{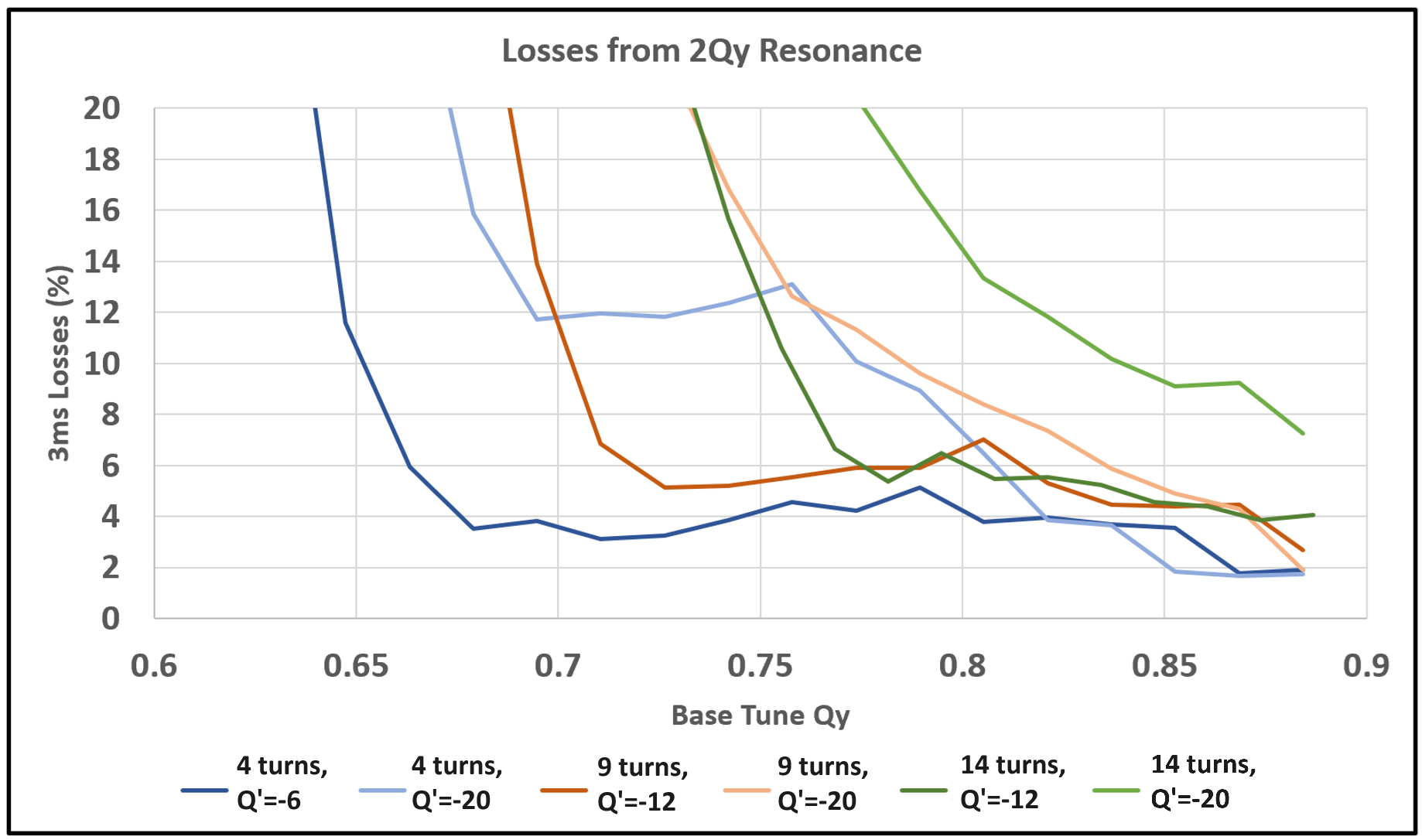}}
    \caption{(upper graph) 95\%  Horizontal emittance at extraction from the Booster and (lower graph) Beam loss after 3~ms, as a function of intensity (increasing with blue, orange, green) and chromaticity (stronger dark to light). \label{fig:2QY_Range}}
\end{figure}

\section{2Qy Resonance Correction}
\label{sec:2qy}

\subsection{Previous work on 2Qy=13 Correction}
\label{sec:2qy_intro}

In the Fermilab Booster, the bare vertical tune is set just below the $Q_{y}=7$ resonance to allow the greatest possible tune-depression before driving emittance growth and losses as a result of the ${2Q_{y}=13}$ resonance. Figure~\ref{fig:2QY_Range} shows the beam losses and emittance as a result of bare vertical tune, intensity, and chromaticity. At the highest intensities, no tuning range remains and performance gains must therefore come from compensating the ${2Q_{y}=13}$ resonance.

\begin{figure}
  \centering
    \includegraphics[width=.6\linewidth]{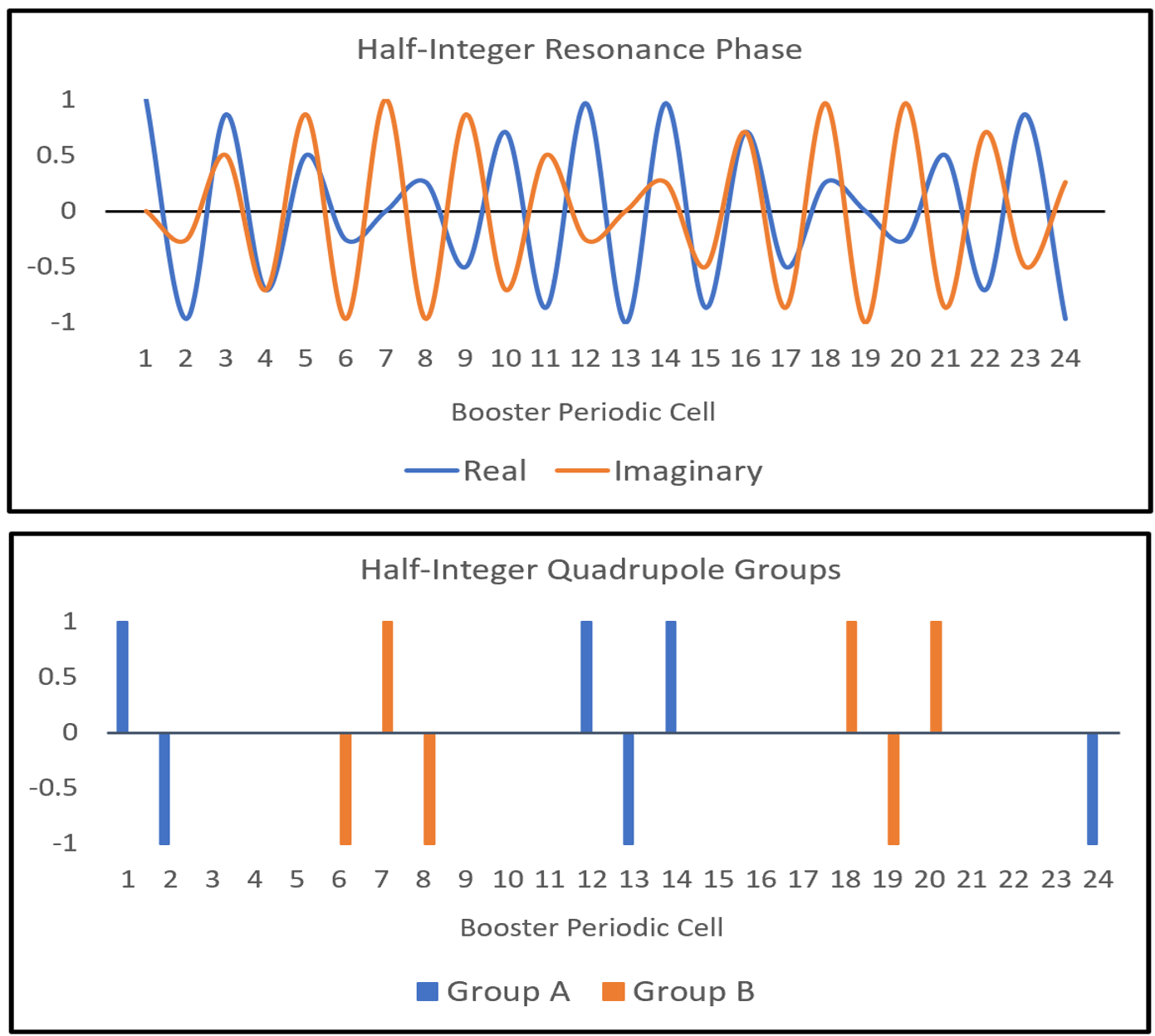}
    \caption{The relative contribution of a quadrupole in each Booster cell to the real (blue line) and imaginary (orange line) parts of the ${2Q_{y}=13}$ resonance. The six quadrupoles which make the strongest contribution to the real and imaginary parts are selected for $2Q_{y}$ cancellation circuits.
    \label{fig:2QyCorr}}
\end{figure}

In each of the 24 Booster cells, there is a quadrupole corrector located at a location with high $\beta_{y}$ and low $\beta_{x}$. Each quadrupole corrector contributes to the ${2Q_{y}=13}$ with a phase-difference of $\pi +\pi/12$ per Booster cell. Figure~\ref{fig:2QyCorr} shows the relative contribution of each quadrupole to the real and imaginary parts of the ${2Q_{y}=13}$. To efficiently cancel the ${2Q_{y}=13}$ resonances, two groups of six quadrupoles are selected to make orthogonal contributions to the resonance. Within each group of six, there are three adjacent quadrupoles alternating in polarity, and another three separated by 12 cells. Table~\ref{tab:2Qy13} lists the specific cells and polarity for group ``A'' and ``B''.

\begin{table}[htbp]
\begin{center}
{\begin{tabular}{|l||c|c|c||c|c|c||}
\hline
Corrector & \multicolumn{3}{c||}{First Three} & \multicolumn{3}{c|}{Second Three} \\
\hline
A$_{2QY=13}$ & \textcolor{Red}{dQL$_{24}$=-A} & \textcolor{ForestGreen}{dQL$_{01}$=+A} & \textcolor{Red}{dQL$_{02}$= -A} & \textcolor{ForestGreen}{dQL$_{12}$=+A} & \textcolor{Red}{dQL$_{13}$= -A}  & \textcolor{ForestGreen}{dQL$_{14}$=+A} \\
B$_{2QY=13}$ & \textcolor{Red}{dQL$_{06}$=-B} & \textcolor{ForestGreen}{dQL$_{07}$=+B} & \textcolor{Red}{dQL$_{08}$= -B} & \textcolor{ForestGreen}{dQL$_{18}$=+B} & \textcolor{Red}{dQL$_{19}$= -B} & \textcolor{ForestGreen}{dQL$_{20}$=+B} \\
\hline
\end{tabular}}
\caption{Two groups of six quadrupoles, termed A \& B, are used to make orthogonal contributions to the $2Q_{y}=13$ resonance. Green text indicates positive polarity and red text indicates negative polarity.}
\label{tab:2Qy13}
\end{center}
\end{table}

\begin{figure}
  \centering
  \subcaptionbox{Booster efficiency as a function of the current setting of two groups of ${2Q_{y}=13}$ compensating quadrupoles. \label{fig:2QyScan}}
    {\includegraphics[width=.38\linewidth]{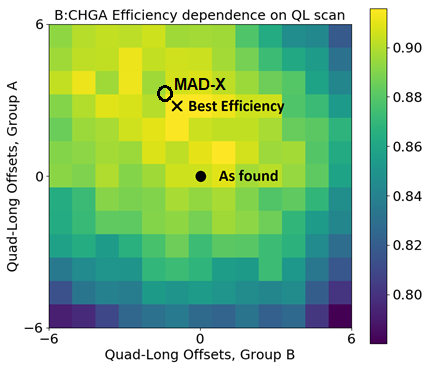}}\qquad
  \subcaptionbox{Low-intensity tunescan for nominal corrector settings and for efficiency-optimized corrector settings (A=1,B=-3). \label{fig:2QyTunescan}}
    {\includegraphics[width=.42\linewidth]{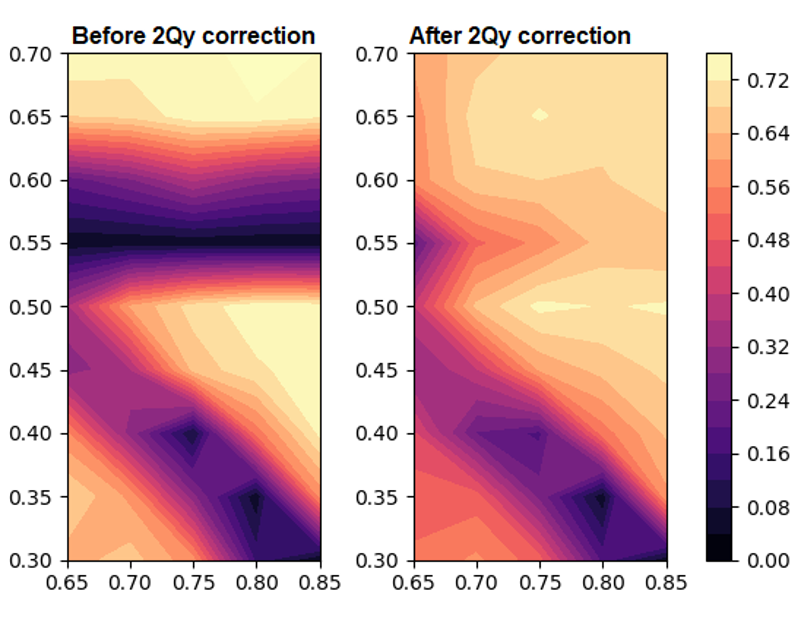}}
    \caption{Use of ${2Q_{y}=13}$ compensating quadrupoles improves Booster efficiency, especially near resonance. \label{fig:2Qy}}
\end{figure}

The settings of corrector groups A and B are scanned to optimize Booster performance, and Figure~\ref{fig:2Qy} shows the impact of the scanned values on efficiency. Booster efficiency is calculated as the ratio of the beam stored at extraction (30~ms into cycle) relative to injection (1~ms into cycle) as measured by the toroid. Traditionally this technique is used with the bare tune moved closed to the resonance, however the scan shown in Figure~\ref{fig:2QyScan} was conducted with nominal bare tunes with an intensity of 4.5e12 protons (so that particles with depressed tunes interact with the resonance). 

The best ${2Q_{y}=13}$ compensation settings were found to be with group A at 3~amps and group B at -1~amp, near the predicted value from a MAD-X tracking study as shown in Figure~\ref{fig:2QyScan}. The MAD-X tracking study was performed by using the 48 quadrupole correctors in two families to move the vertical tune to 6.53, and then varying the ${2Q_{y}=13}$ compensation settings to minimize the wave in $\beta_{y}$ as a result of the resonance.

Figure~\ref{fig:2QyTunescan} shows a tunescan through $Q_{y}=6.5$ with nominal settings and with best ${2Q_{y}=13}$ compensation settings. Although the ${2Q_{y}=13}$ line is not completely corrected, it is clear that the compensation circuits are effectively targeting the ${2Q_{y}=13}$ resonance.

Figure~\ref{fig:2QyScan} shows an $\sim 1\%$ improvement in beam efficiency from $2Q_{y}=13$ compensation. Unfortunately, the compensation settings also caused an overall shift of losses from the collimators to the more sensitive extraction section. 

\begin{figure}[H]
  \centering
    \includegraphics[width=.6\linewidth]{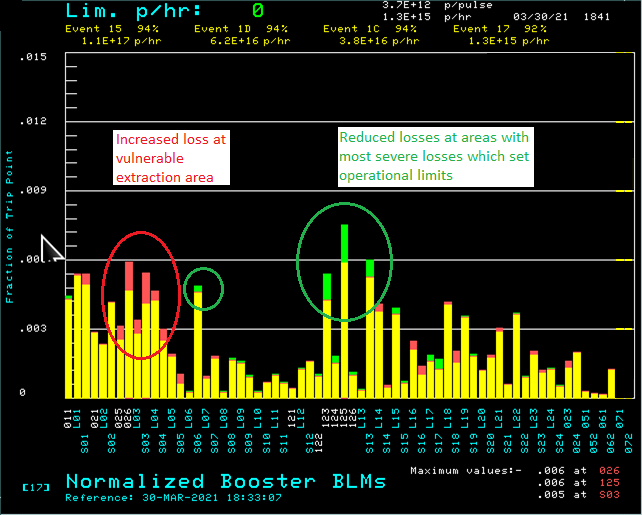}
    \caption{Each bar represents a different loss monitor, as a fraction of its trip point. The red represents where the losses in the $2Q_{y}$ compensations are worse than operational settings and the green represents where they are improved. \label{fig:2QyLoss}}
\end{figure}

Figure~\ref{fig:2QyLoss} shows the change in loss monitors around the ring, each normalized to their trip point. The $2Q_{y}$ compensation reduces losses in the collimation location early in the cycle when space-charge forced dominate and increases losses in the extraction locations later in the cycle. Further studies confirmed these loss patterns were not a result of orbit changes or quad-steering effects.

\begin{figure}[H]
  \centering
    \includegraphics[width=.6\linewidth]{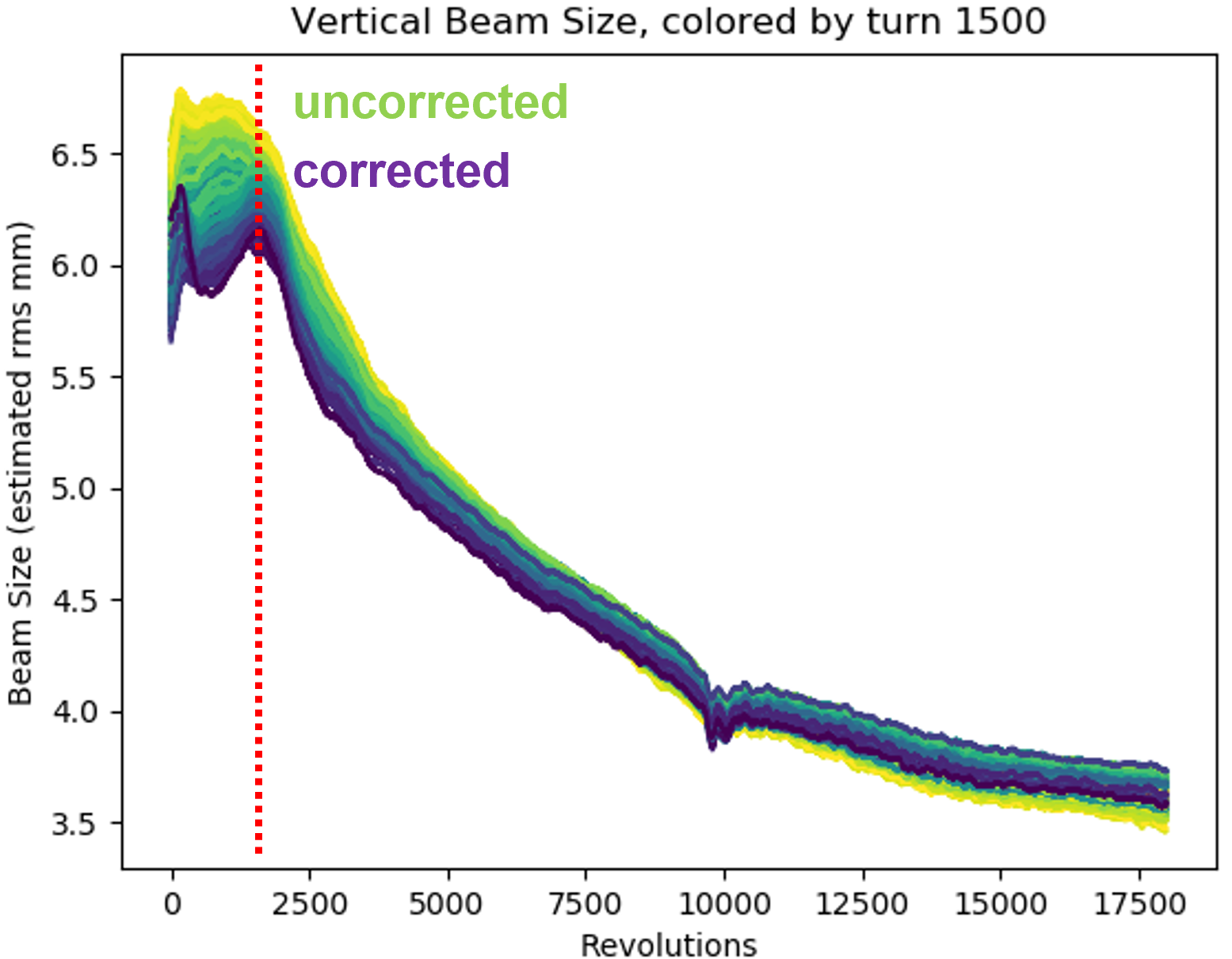}
    \caption{Size of vertical beam profile measured by Booster IPM, for several cycles making up a ${2Q_{y}=13}$ compensation scan. Each line is colored by size on turn 1500 (indicated by red line), which is changed by extraction. \label{fig:2QyVsize}}
\end{figure}

The IPM data indicates that the vertical beam size at turn 1500, is also minimized for the ${2Q_{y}=13}$ compensation settings which optimize Booster efficiency. Figure~\ref{fig:2QyVsize} shows how the vertical beam size progresses for each value in a ${2Q_{y}=13}$ correction scan. By the end of the cycle, the uncorrected beam experiences less vertical emittance growth than a comparable beam with ${2Q_{y}=13}$ compensation. Further studies have confirmed that this is not a scraping effect, and could not be recreated by other loss mechanisms. 

We have considered two possible explanations offered for the paradoxical result that $2Q_{y}$ correction is associated with worse performance later in the cycle. The first possible explanation is that the quad-correction had an unintended effect on other particles resonances. Follow-up studies indicate that the horizontal half-integer resonance ${2Q_{x}=13}$ was not impacted, and a study of ${2Q_{y}=14}$ is presented in Section~\ref{sec:2qy_14}. A preliminary study of higher-order resonances is presented in Section~\ref{sec:2qy_tune}.

\begin{figure}
  \centering
    \includegraphics[width=.6\linewidth]{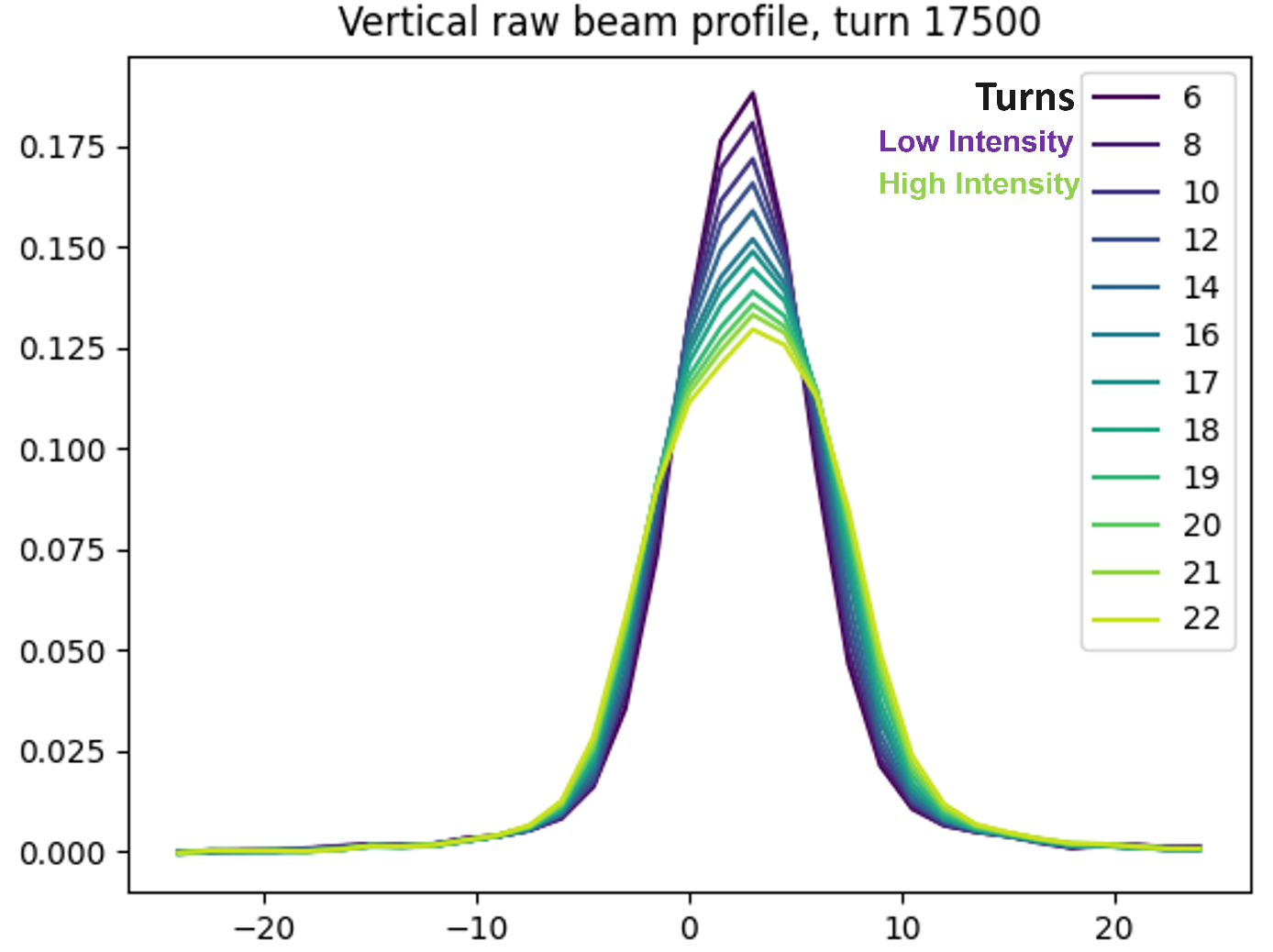}
    \caption{Vertical beam profile measured by Booster IPM as a function of intensity. The profiles have been rescaled (along the horizontal axis) to have the same RMS value. \label{fig:2QyShape}}
\end{figure}

The second possible explanation is that half-integer itself has a protective effect on the beam performance. The $2Q_{y}=13$ resonance approached from above has the effect of not only blowing up the beam emittance, but specifically hollowing out the core of the beam. Figure~\ref{fig:2QyShape} shows the change in the IPM beam profile as a function of intensity. A broader beam profile may be less susceptible to beam instabilities than a beam with comparable emittance and intensity but a denser core (as a result of $2Q_{y}=13$ correction). Further studies found that changing the chromaticity by $\pm 4$ did not have any effect on the effectiveness of the $2Q_{y}=13$ compensation. A preliminary attempt to modify the early beam emittance with a transverse anti-damper was not successful.

\subsection{Role of 2Qy=14}
\label{sec:2qy_14}

Although in 2019 the $2Q_{y}$ correction could be optimized from nominal tunes at high intensities, the same procedure had prohibitive signal-to-noise ratio in 2021 and thereafter. The reason is that the operational vertical tune was raised from $Q_{y}\approx6.84$ to $Q_{y}\approx6.88$ to better accommodate high-intensity operations.

\begin{figure}
  \centering
    {\includegraphics[width=.3\linewidth]{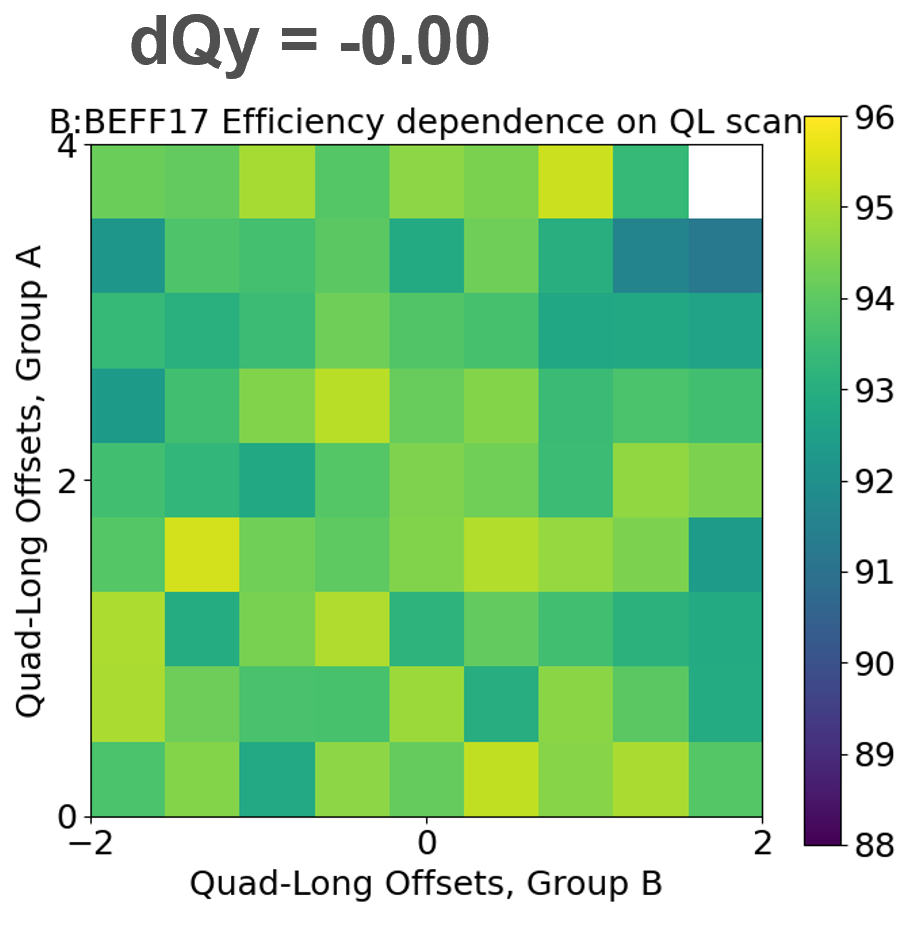}}\quad
    {\includegraphics[width=.3\linewidth]{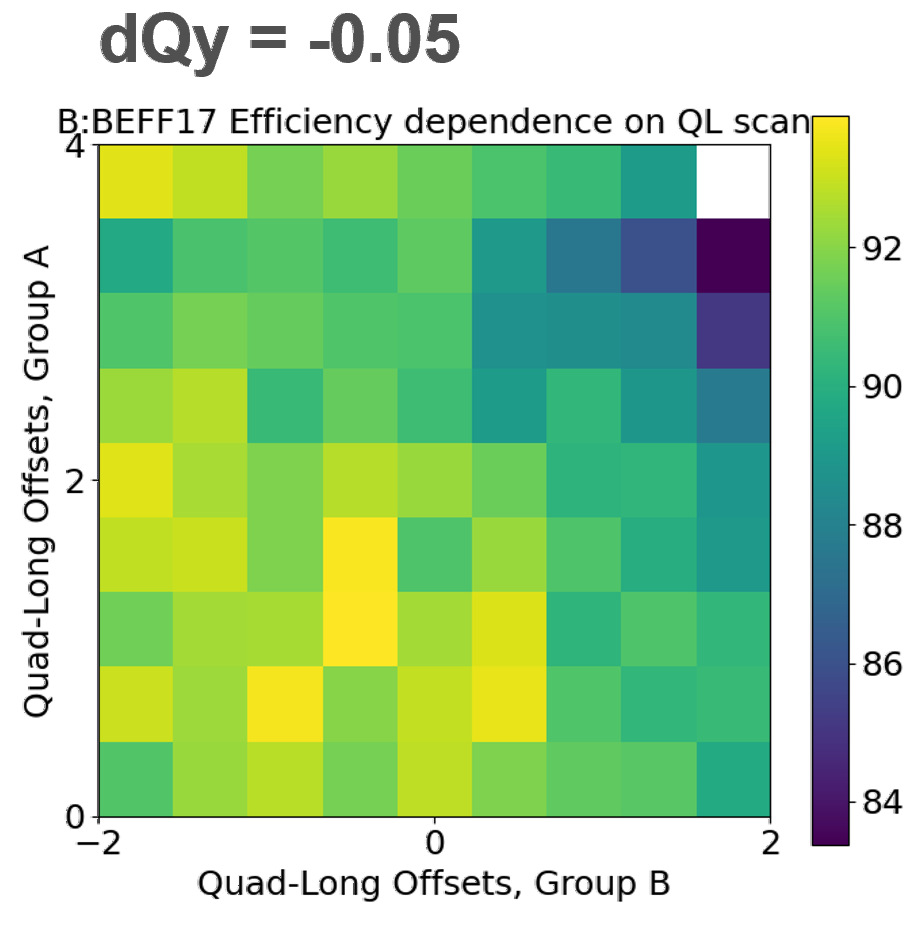}}\quad
    {\includegraphics[width=.3\linewidth]{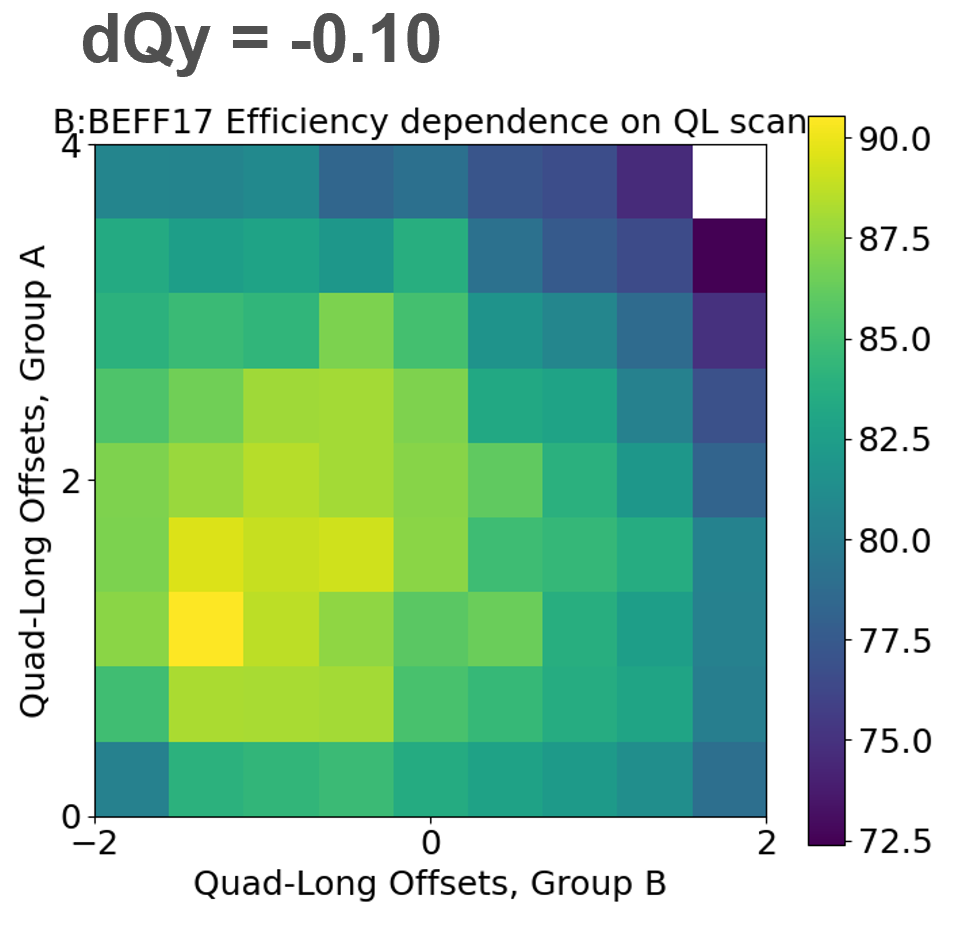}}
    \caption{Booster efficiency as a function of the current setting of two groups of $2Q_{y}=13$ compensating quadrupoles. Corrector Scan shown for nominal tune settings (left), bare vertical tune lowered by -0.05 (center), and bare vertical tune lowered by -0.1 (right).
    \label{fig:2Qy_dQy}}
\end{figure}

When a mild tune depression was introduced (using all quads in short and long sections as two families), it became possible again to ascertain the best correctors settings for mitigation of the $2Q_{y}$ resonance. In Figure~\ref{fig:2Qy_dQy} we see Booster efficiency against 2Qy=13 compensation corrector settings as a function of tune depression.

We can compare Figure~\ref{fig:2Qy_dQy} from July 2023 to Figure~\ref{fig:2QyScan} from February 2020 (noting the change in axis labels). The best 2$Q_{y}$ correction is also changed as a result of drift in regular Booster tuning parameters.

\begin{figure}
  \centering
    \includegraphics[width=.6\linewidth]{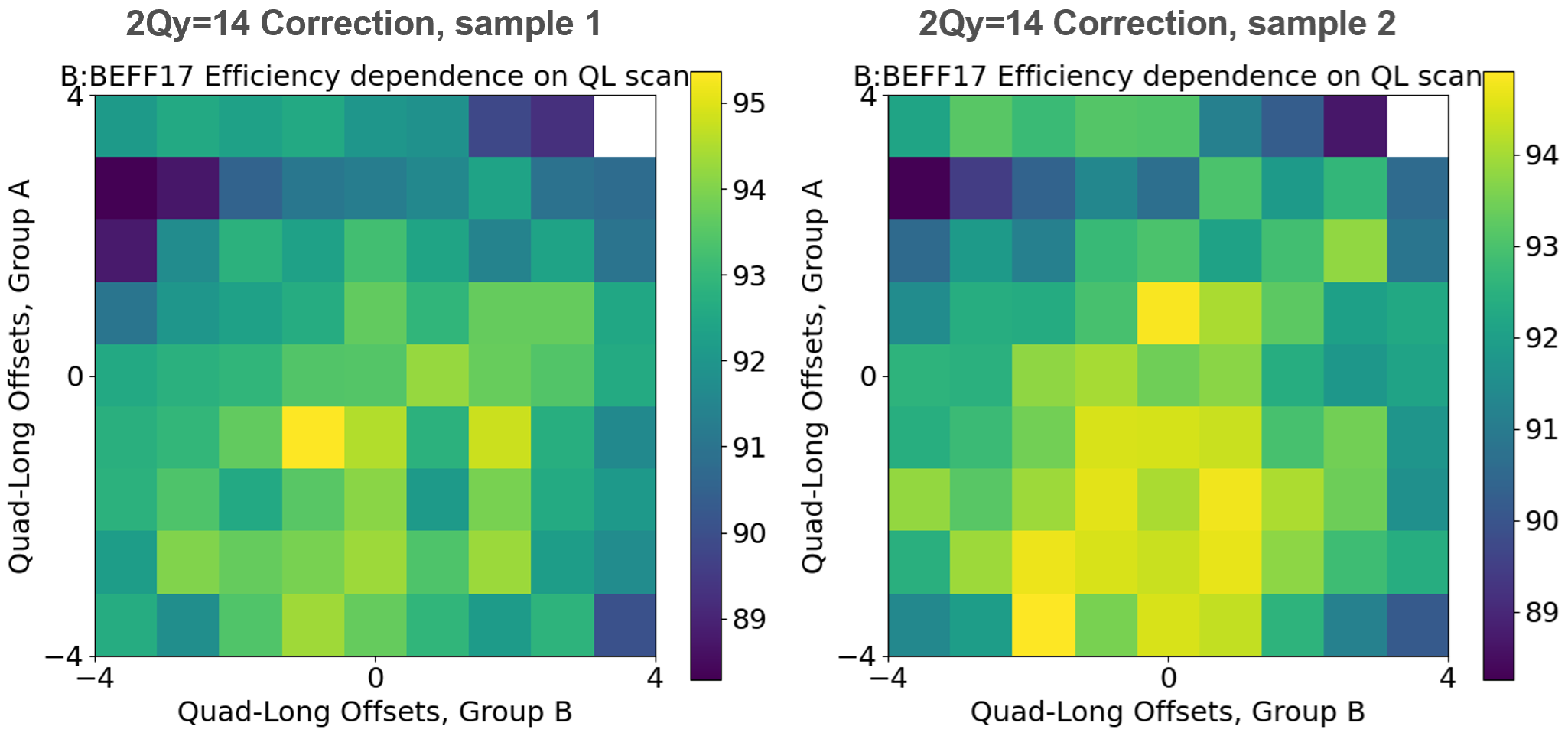}
    \caption{Booster efficiency as a function of the current setting of two groups of ${2Q_{y}=14}$ compensating quadrupoles. There is a large cycle-to-cycle variation, and two samples taken under identical machines settings are shown. \label{fig:2Qy14}}
\end{figure}

With the higher vertical tune in operations, it raises the question as to whether the ${2Q_{y}=14}$ resonance should also have an impact on beam performance. A new set of correctors ``A'' and ``B'' were identified to target ${2Q_{y}=14}$ resonance, shown in Table~\ref{tab:2Qy14}. Figure~\ref{fig:2Qy14} shows two separate ${2Q_{y}=14}$ corrector scans under identical conditions, which indicate a possible 0-1\% improvement in beam efficiency by optimizing the ${2Q_{y}=14}$ correction to A$_{2QY=14}\approx-3$, B$_{2QY=14}\approx-1$

\begin{table}[htbp]
\begin{center}
{\begin{tabular}{|l||c|c|c||c|c|c||}
\hline
Corrector & \multicolumn{3}{c||}{First Three} & \multicolumn{3}{c|}{Second Three} \\
\hline
A$_{2QY=14}$ & \textcolor{Red}{dQL$_{24}$=-A} & \textcolor{ForestGreen}{dQL$_{01}$=+2A} & \textcolor{Red}{dQL$_{02}$= -A} & \textcolor{Red}{dQL$_{12}$=-A} & \textcolor{ForestGreen}{dQL$_{13}$= +2A}  & \textcolor{Red}{dQL$_{14}$==A} \\
B$_{2QY=14}$ & \textcolor{Red}{dQL$_{09}$=-B} & \textcolor{ForestGreen}{dQL$_{10}$=+2B} & \textcolor{Red}{dQL$_{11}$= -B} & \textcolor{Red}{dQL$_{21}$=-B} & \textcolor{ForestGreen}{dQL$_{22}$=+2B} & \textcolor{Red}{dQL$_{23}$= -B} \\
\hline
\end{tabular}}
\caption{Two groups of six quadrupoles, termed A \& B, are used to make orthogonal contributions to the ${2Q_{y}=14}$ resonance. Green text indicates positive polarity and red text indicates negative polarity.}
\label{tab:2Qy14}
\end{center}
\end{table}

We predicted that chromaticity was responsible for particle loss below the ${2Q_{y}=14}$ resonance, and designed a series of beam studies to investigate that possibility. ${2Q_{y}=14}$ correction scans were performance at 17 turns intensity, with six samples with chromaticities from $Q_{y}^{\prime}=-10$ to $Q_{y}^{\prime}=-2$ and again at 12 turns intensity, with seven samples between $Q_{y}^{\prime}=-10$ and $Q_{y}^{\prime}=-2$. In some samples there appears to be a preference for correction in the same quadrant (A and B negative), and others the signals appear to be complete noise. Notably, there was no indication that the sensitivity to the ${2Q_{y}=14}$ correction depended on intensity or chromaticity. Follow-up studies have to be conducted with greater statistics to investigate the value of ${2Q_{y}=14}$ correction and the origin of its influence on the beam (if any).

\subsection{Impact of Intensity \& Tune on Emittance Growth}
\label{sec:2qy_tune}

Recent work by Asvesta at the CERN~PS~\cite{Asvesta20} examines the complex interplay between optics correction and space-charge induced resonances. Our upcoming work will analyze the space-charge resonance driving terms (RDTs), parameterized as a function of several adjustments in beam optics. The presence of higher-order space-charge induced resonances might also be able to inferred empirically by varying the intensity, bare vertical tune, and observing the emittance growth as the beam ramps (and the space-charge tune-spread shrinks).

IPM data were collected for five intensity and five tune conditions making for 25 cases. The five intensity conditions were 6 turns, 10 turns, 14 turns, 18 turns, and 20 turns - corresponding to injected beam intensities of 1.7e12, 2.8e12, 4.0e12, 5.1e12, and 5.7e12 respectively. The five tune conditions were vertical tune changes of $dQ_{y}=-0.0$, $dQ_{y}=-0.05$, $dQ_{y}=-0.1$, $dQ_{y}=-0.15$, $dQ_{y}=-0.2$ corresponding to tunes of $Q_{y}\approx6.88$, $Q_{y}\approx6.83$, $Q_{y}\approx6.78$, $Q_{y}\approx6.73$, $Q_{y}\approx6.68$ respectively. Due to cycle-to-cycle variation in beam quality, at least ten cycles of data were collected for each of the 25 cases.

\begin{figure}[H]
  \centering
  \subcaptionbox{Vertical beam size as measured by IPMs, nominal tunes, at several intensities. \label{fig:VertSizeTune}}
    {\includegraphics[width=.45\linewidth]{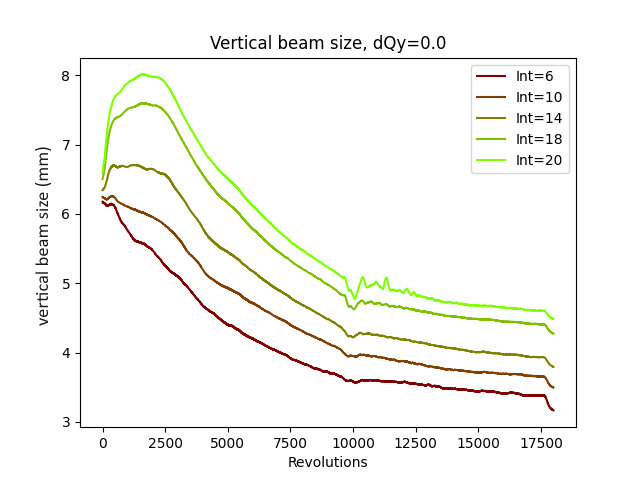}}\qquad
  \subcaptionbox{RMS vertical emittance calculated and calibrated from IPMs, nominal tunes, at several intensities. \label{fig:VertEmitTune}}
    {\includegraphics[width=.45\linewidth]{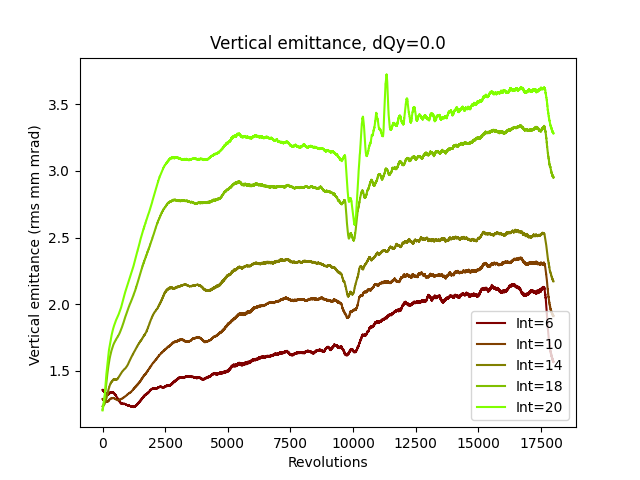}}
    \caption{Vertical beam size and emittance, for several beam intensities. \label{fig:Vert_SzEm_Tune}}
\end{figure}

Figure~\ref{fig:VertSizeTune} and Figure~\ref{fig:VertEmitTune} show the vertical beam size and calculated emittance measured by the IPMs, for nominal tune for each intensity case. Each case is averaged across all available samples and the vertical emittance calculation incorporates the IPM calibration from \cite{ShiltsevIPM}. There are some artifacts in the IPM signal for transition-crossing and orbit changes (similar to \cite{ShiltsevEldred}, the predominant change with intensity is the magnitude and duration of the initial emittance blow-up.

\begin{figure}[H]
  \centering
  \subcaptionbox{Vertical beam size as measured by IPMs, at 10 turns (4.0e12 intensity), at several tunes. \label{fig:VertSizeInt}}
    {\includegraphics[width=.45\linewidth]{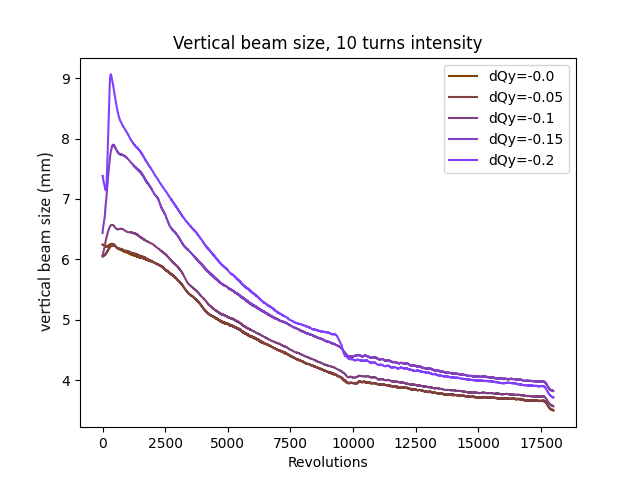}}\qquad
  \subcaptionbox{RMS vertical emittance calculated and calibrated from IPMs, at 10 turns (4.0e12 intensity), at several tunes.  \label{fig:VertEmitInt}}
    {\includegraphics[width=.45\linewidth]{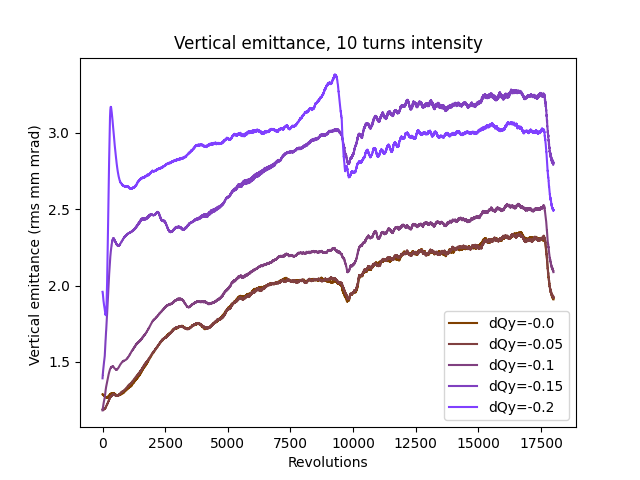}}
    \caption{Vertical beam size and emittance, for several vertical tunes (expressed as change from nominal). \label{fig:Vert_SzEm_Int}}
\end{figure}

Figure~\ref{fig:VertSizeInt} and Figure~\ref{fig:VertEmitInt} show the vertical beam size and calculated emittance measured by the IPMs, for 4.0e12 injected particles and for each tune case. Each case is averaged and calibrated. Moving the tune closer to $2Q_{y}=13$ resonance causes a similar emittance blow-up to increasing the charge. Each case shows a feature (also present in Figure~\ref{fig:Vert_SzEm_Tune}) in which vertical emittance dips or plateaus roughly between turn 3500 and 5000. If this feature was associated with a specific space-charge depressed tune, it would be occurred later in the cycles with reduced vertical tunes (which is not the case). In preliminary analysis the emittance-growth in the 25 cases are not consistent with any specific higher-order resonance impacting the Booster at depressed tunes above the $2Q_{y}=13$ resonance.

\section{Mode Coupling Instabilities with Strong Space-charge}
\label{sec:headtail}

\subsection{Background \& Prior Work}

Transverse Mode Coupling Instability (TMCI) is a well documented beam instability present in synchrotrons with significant impedance~\cite{Chao93,Chao23}. Similar to the beam break up (BBU) in linacs, transverse wakes from the head of the bunch induce kicks at the tail of the bunch which resonate with the betatron frequency. Unlike BBU however, the beam undergoes synchroton motion which interchange the particles in the head and tail of the bunch. The beam becomes unstable when wake effects cause the syncho-betatron sidebands to couple with each other.


In 1998, Blaskiewicz~\cite{Blaskiewicz98} incorporated a simple space-charge model into the analysis of mode coupling instabilities, appearing to show that space-charge had the effect of mitigating the TMCI instability. In fact it would later be shown that only an intermediate space-charge mitigates the beam instability~\cite{Buffat21,Balcewicz23}.

The interaction of mode coupling instabilities with strong space-charge is quite complex, and the subject of active theoretical and experimental research. In 2009, Burov developed a formalism for head-tail modes in beams with strong space-charge~\cite{Burov09}, which was simulated in 2015 by Macridin~et.~al.~\cite{Macridin15}. Ultimately consideration of mode coupling instabilities with strong space-charge led Burov to predict a new ``convective'' beam instability~\cite{Burov19}.

Whereas TMCI shows exponential growth in time of the oscillation amplitude, convective motion is unstable across the length of the bunch - the instability shows a progressing ``amplification'' of the oscillation amplitude from head to tail. If there is a synchrotron oscillation, the head-tail patterns persists with the particles currently at the tail oscillating at the larger amplitude than the particles currently at the head. Convective instabilities can be further sorted into saturating and absolute, where in the latter case the beam is also unstable across time.

Burov's work was motivated in part by the need to understand the behavior of mode-coupling instabilities in the CERN SPS. As part of the CERN LIU project~\cite{Damerau:1976692}, a contemporary outlook on the SPS mode-coupling instabilities is given in \cite{Buffat22}. This recent work confirms that the TMCI models of the CERN SPS must incorporate space-charge effects to give accurate results. Recently the presence of TMCI and convective motion has also been studied in the Fermilab Recycler using long bunches~\cite{Balcewicz23,Mohsen23}.

The Fermilab Booster is vulnerable to mode-coupling instabilities with strong space-charge during transition crossing, because the phase-slip factor and chromaticity approach zero simultaneously. During the 2019 International Fermilab Booster studies~\cite{EldredCapstone}, a ``zero-chromaticity transition'' (ZCT) ramp was prepared to better study the instability. In the ZCT ramp, the sextupole were tuned to keep chromaticity weak (less than 2 units) for 6ms in the vicinity of transition. Under the ZCT condition, the Fermilab Booster was found to suffer catastrophic $\sim$20\% beam loss at moderate intensities of $\sim$3.5~e12 protons. The loss was concentrated in the tail of each bunch, and the tail featured ten times larger amplitude of oscillation than the head. Each bunch became unstable independently, within a $\sim$100 turn window near transition. The 2019 International Fermilab Booster study found, the instability takes place at $\sim$3.5e12 under ZCT conditions, while the 2023 study presented here finds the instability takes place at $\sim$5.5e12 under nominal conditions. Understanding and mitigating this beam instability will be necessary for PIP-II operations.

\subsection{Threshold of Mode Coupling Instability \& Implications for PIP-II Era Operation}

Recall that Figure~\ref{fig:Instability_Elog} shows a high-intensity beam instability near transition. The nature of this beam instability will be measured using a high-bandwidth transverse pickup and must also be analyzed in light of the chromaticity changes near transition.

\begin{figure}[htbp]
\centering
\includegraphics[width=.6\textwidth]{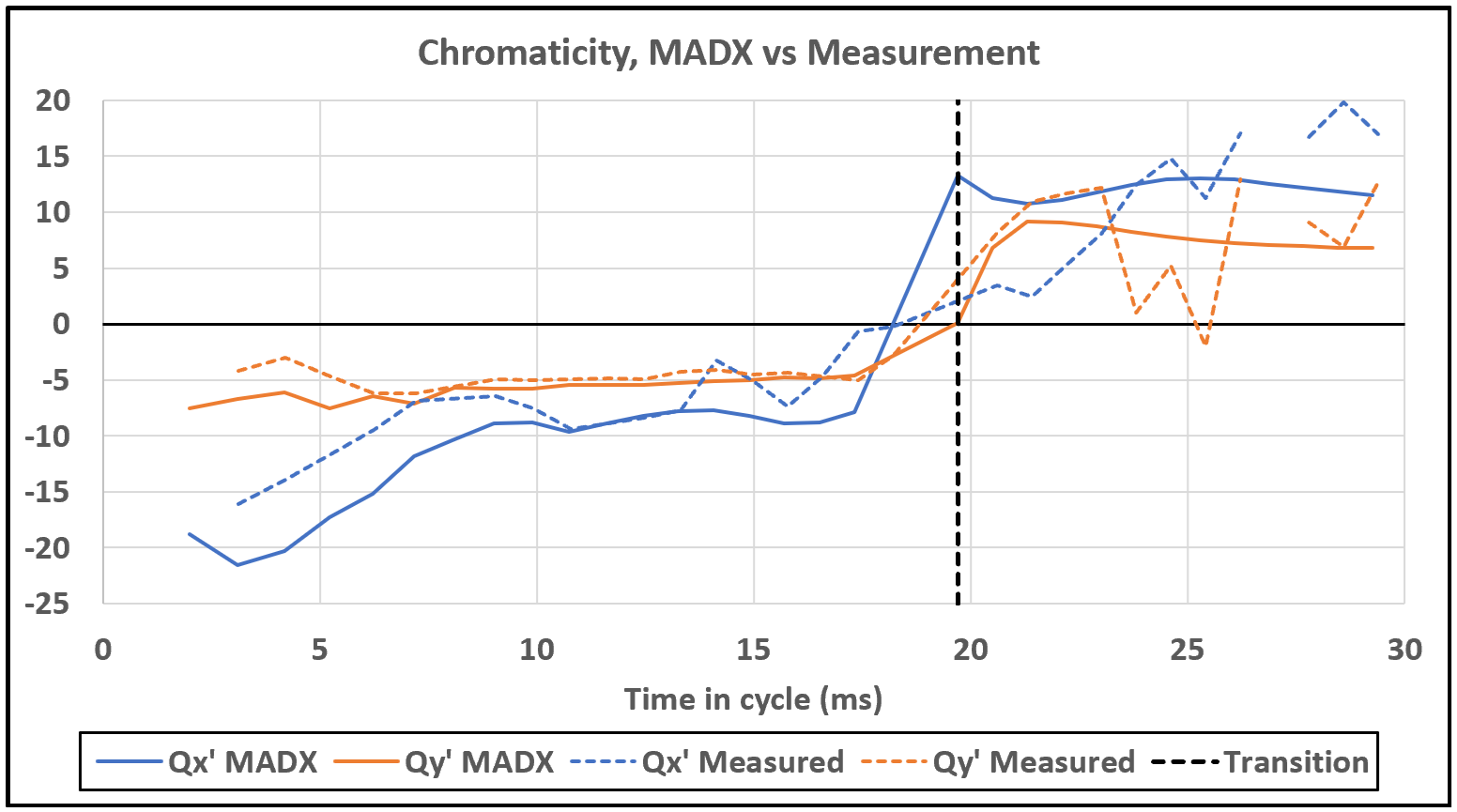}
\caption{Comparison of MAD-X model and measured values of chromaticity throughout the ramp.}
\label{fig:Chromaticity}
\end{figure}

Consistent with the Booster's TMCI and coupled bunch instability (CBI) mitigation strategy, chromaticities are designed to be strongly negative at low-energies, moderately negative below transition, and moderately positive after transition. Figure~\ref{fig:Chromaticity} shows the MADX model and machine measurements of the chromaticity through out the ramp.

For a betatron tune measurement, the beam is kicked or ``pinged'' every 1000 revolutions, and after each kick the tunes are measured by 48 turn-by-turn BPMs. Additionally the beam momentum is varied by changing the radial feedback of the RF system (i.e. inferring the momentum change from the corresponding dispersive effect). Consequently the chromaticity measurement is calculated by the change in tune measurements divided by the change in fractional momentum $\delta$. The sextupole components of the Booster gradient dipoles are themselves a fourth-order polynomial fit to 2015 chromaticity measurements. The model and measurements are in agreement that the vertical chromaticity switches from negative to positive at transition, whereas the MAD-X model suggests the horizontal chromaticity is positive at transition.

The transverse bunch-by-bunch dampers system uses a high-bandwidth stripline pickup. The stripline has been developed into a diagnostic using an RF hybrid to produce sum and difference signals, horizontally and vertically. Then with 4 GHz oscilloscope, a dipole measurement of each $\sim$6~ns bunch is possible at a 0.25~ns resolution, for all 81 bunches in a revolution and for up to about 2500 revolutions at a time.

The RF hybrid system has the capability of using a variable gain for each plate to remove orbit effects from the measurement of betatron and intrabunch motion, although this feature was not used for this study. The instrumentation can also support measurements of horizontal and vertical motion simultaneously, although for this study only the vertical motion was used (previous studies established the head-tail motion is predominately vertical).

\begin{figure}
  \centering
  \subcaptionbox{Sum signal (blue) and difference signal (orange) from stripline in 4ms window around transition. Longitudinal quadrupole oscillations are observed after the bunch-length minimum corresponding to transition. \label{fig:Stripline_Raw}}
    {\includegraphics[width=.4\linewidth]{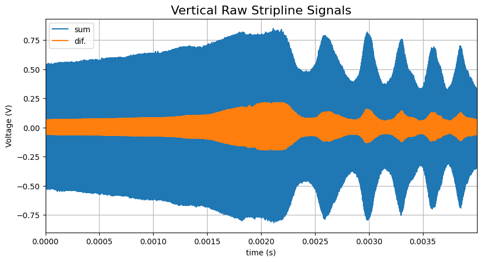}}\qquad
  \subcaptionbox{Integrated sum signal (solid black) and difference signal (dotted) for a single bunch. After subtracting off a linear pedestal, the position within the bunch is calculated as the ratio (blue) scaled by a calibration factor. \label{fig:Stripline_Int}}
    {\includegraphics[width=.4\linewidth]{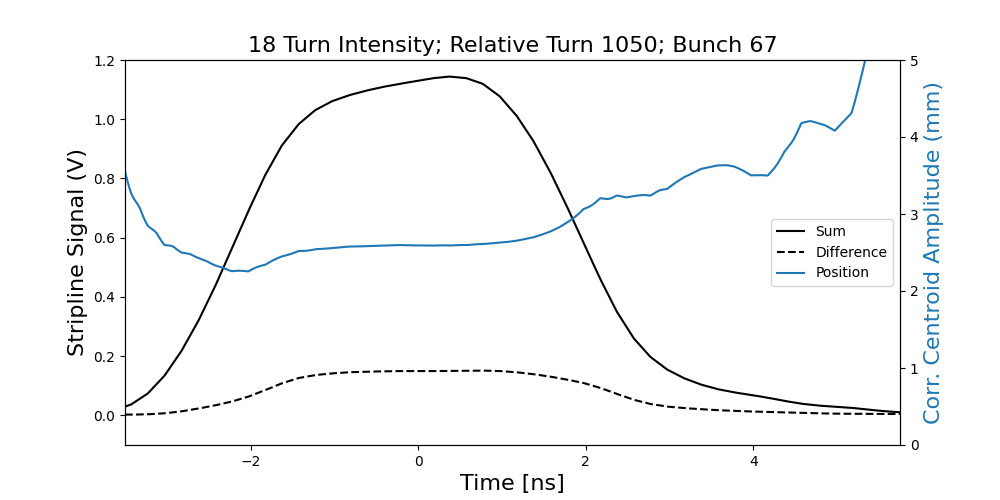}}
    \caption{Processing of raw stripline signals into measurement of position within bunch. \label{fig:Stripline}}
\end{figure}

Figure~\ref{fig:Stripline_Raw} shows the raw sum \& difference signals from the RF hybrid stripline measurement. The raw signal is obtained from the change in beam current, and therefore must be integrated to obtain line charge (integrated sum signal) and dipole moment (integrated difference signal). Figure~\ref{fig:Stripline_Int} shows the integrated signals, with a linear pedestal subtracted from each bunch. The position signal is proportional to the ratio of the integrated difference signal to the integrated sum signal. This position signal from the stripline was cross-calibrated with BPMs, indicating that these arbitrary units should be multiplied by 19.88~mm to convert to physical units.

\begin{figure}
  \centering
  \subcaptionbox{Integrated stripline signals below instability threshold. \label{fig:SumDiff_1}}
    {\includegraphics[width=.4\linewidth]{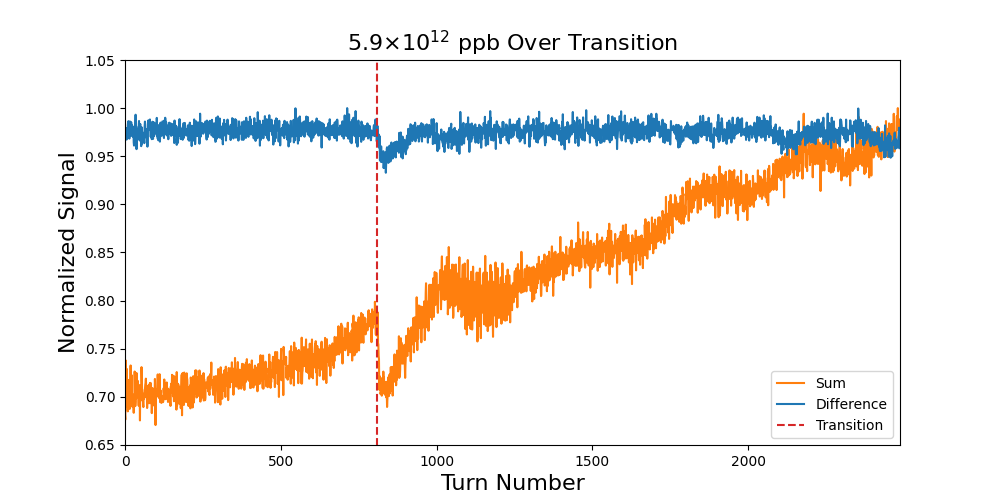}}\qquad
  \subcaptionbox{Integrated stripline signals above instability threshold. \label{fig:SumDiff_2}}
    {\includegraphics[width=.4\linewidth]{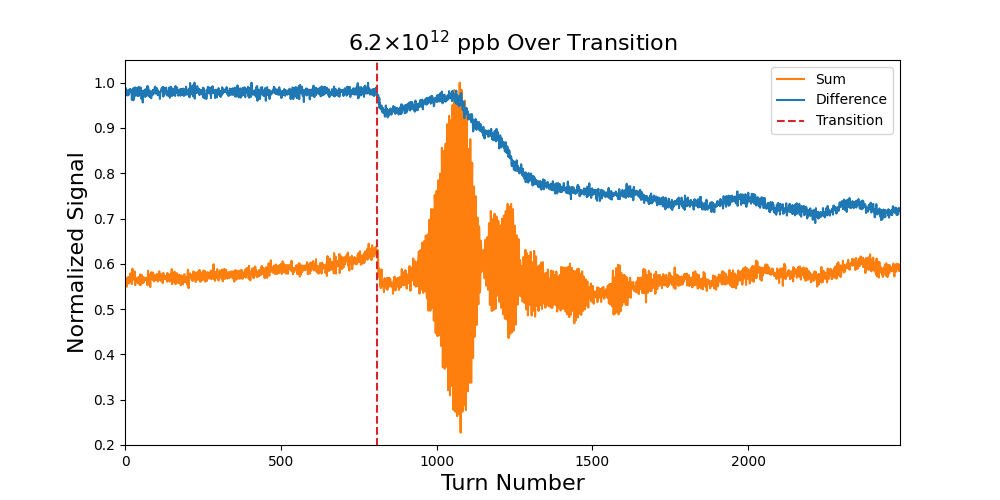}}
    \caption{Integrated sum signal (orange) and integrated difference signal (blue) for below and above instability threshold. Red dashed line indicates transition. \label{fig:SumDiff}}
\end{figure}

Figure~\ref{fig:SumDiff} shows the line charge and dipole moment  (integrated sum and difference) signals in the vicinity of transition below and above the instability threshold. In Figure~\ref{fig:SumDiff_1} (below threshold), vertical motion coincides with imperfect transition crossing and the level of the integration sum signal indicates minimal loss. In Figure~\ref{fig:SumDiff_2} (above threshold), the vertical motion around transition seeds the beam instability which grows rapidly until significant beam loss takes place. The beam instability damps as a result of the drop in intensity, restoration of synchrotron motion, and/or increase in chromaticity.

\begin{figure}
  \centering
  \subcaptionbox{Integrated stripline signals below instability threshold, despite kicking the beam. \label{fig:SumDiff_Ping_1}}
    {\includegraphics[width=.4\linewidth]{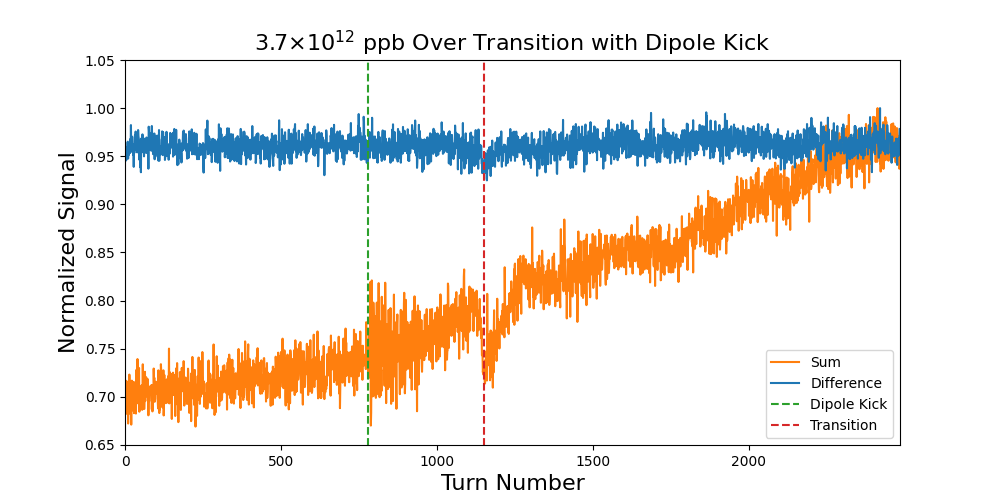}}\qquad
  \subcaptionbox{Integrated stripline signals above instability threshold as a result of kicking the beam. \label{fig:SumDiff_Ping_2}}
    {\includegraphics[width=.4\linewidth]{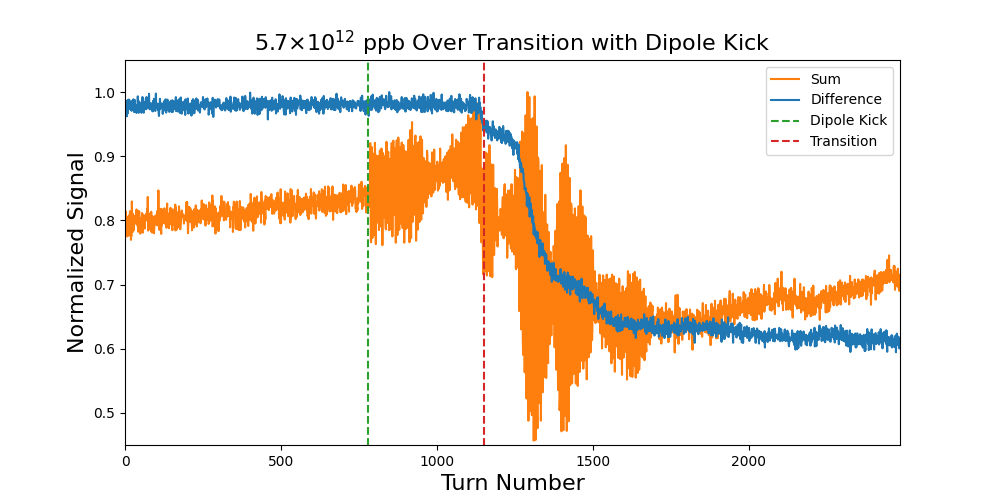}}
    \caption{Integrated sum signal (orange) and integrated difference signal (blue) for below and above instability threshold. Dashed lines indicates dipole kick (green) and transition (red). \label{fig:SumDiff_Ping}}
\end{figure}

A kicker is typically used to ``ping'' the beam as part of the betatron measurement system. For this beam study, a single dipole kick was configured to roughly 400 turns before transition. Figure~\ref{fig:SumDiff_Ping} shows the line charge and dipole moment signals in the vicinity of transition below and above the instability threshold. In Figure~\ref{fig:SumDiff_Ping_1} (below threshold) the dipole kick mostly decoheres in 200 turns, well before transition. In Figure~\ref{fig:SumDiff_Ping_2} (above threshold) the ping seeds the beam instability before transition-crossing and the losses are incurred immediately after transition. Notice that the presence of the dipole kick reduces the intensity threshold necessary for catastrophic beam loss.

\begin{figure}[htbp]
\centering
\includegraphics[width=.6\textwidth]{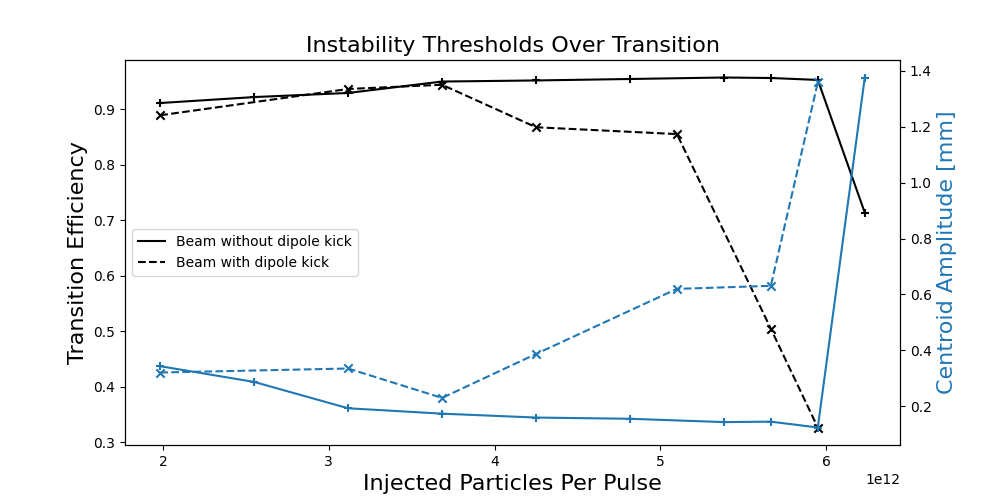}
\caption{Transmission efficiency (black) compared to bunch centroid oscillation amplitude (blue), for kicked (dashed) and nominal (solid) beams.}
\label{fig:Threshold}
\end{figure}

Figure~\ref{fig:Threshold} visualizes the threshold of the beam instability by showing the drop in transmission efficiency and rise in centroid oscillation amplitude with injected beam intensity. For this analysis, the transmission efficiency is estimated from the stripline data as a ratio of the integrated sum signal before and after transition. The centroid amplitude is calculated from the position signal obtained by binning together all data from each bunch. The signals are plotted as a function of injected particle intensity, which is somewhat higher than the number particles at transition (see Figure~\ref{fig:Instability_Elog}). Plotting the transmission against the centroid motion, it is clear the beam losses associated with a beam instability occur above 6e12 particles injected, and that seeding the instability with a kick lowers the instability threshold.

The presence of this fast mode coupling instability with strong space-charge may present a challenge to achieving PIP-II intensity objectives (accelerating 6.5e12 protons through transition with low loss). Our primary mitigation strategy is to develop alternate chromaticity curves, so that the loss of synchrotron motion does not coincide with the loss of chromatic detuning. Preliminary studies show the intensity threshold is raised by a moderate change shift in the chromaticity curve. More aggressive modifications of the chromaticity curve may be complicated by the fact that changes to the sextupole correctors also impact the second-order momentum compaction factor (a critical parameter for transition-crossing). For 24-fold symmetric sextupole correction, the chromaticity and second-order momentum compaction effects generally cannot be decoupled. However by implementing a mild dispersion-wave (traditionally used in a resonant $\gamma_{T}$-jump, e.g. \cite{Teng70}) with existing Booster correctors, the chromaticity can be corrected independently from second-order momentum compaction ($K_{2}D_{X}$ vs $K_{2}D_{x}^{3}$ dependece).

Separately, there is work underway exploring the possibility of implementing a resonant $\gamma_{T}$-jump in the Booster, either by using existing quadrupole correctors, upgrading the correctors, or installing dedicated $\gamma_{T}$-jump magnets. If these efforts are successful, the loss of synchrotron focusing can be avoided and the transverse instabilities will be greatly mitigated.

\subsection{Characterization of Mode Coupling Instability with Strong Space charge}

In Booster operations, only 81 out of the 84 RF buckets in the Booster are occupied with bunches (a 3-bucket extraction gap is created after RF capture). Consequently, each revolution of the beam can be unambiguously identified in data analysis of the line charge and the motion of each bunch can be tracked independently. Tracking the same bunch across many revolutions, we decided to synchronize the data using the center of the bunch rather than the center of the RF phase (although since the beam is nearly isochronous, this is a minor distinction).

\begin{figure}
  \centering
  \subcaptionbox{Stable bunch motion during turns 700-800, when the beam is accelerating. \label{fig:Instability_1}}
    {\includegraphics[width=.4\linewidth]{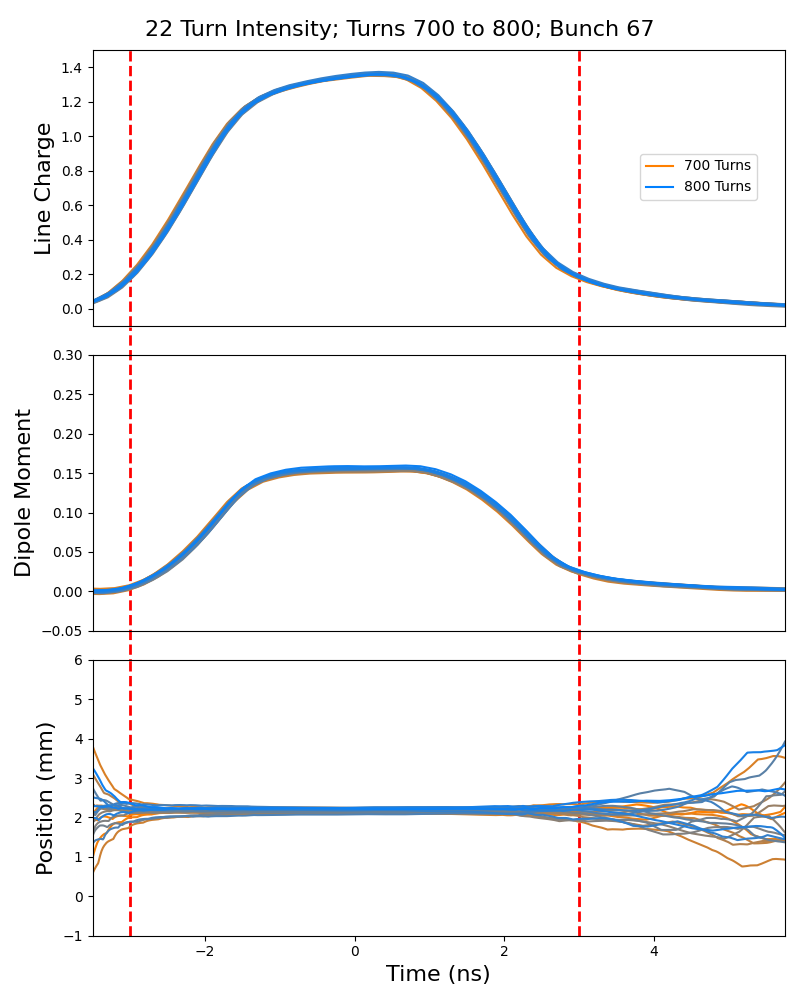}}\qquad
  \subcaptionbox{Tail-dominated motion during turns 900-1150, when the beam approaches transition. \label{fig:Instability_2}}
    {\includegraphics[width=.4\linewidth]{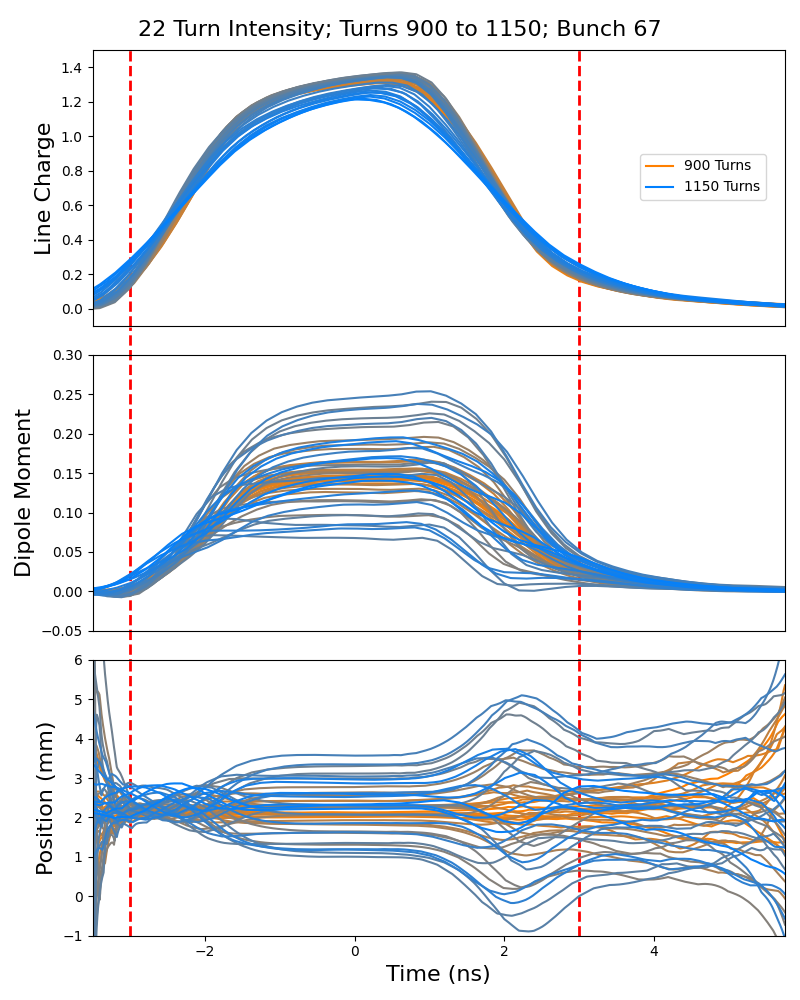}}
  \subcaptionbox{Tail-dominated particle loss during turns 1150-1250, immediately after transition. \label{fig:Instability_3}}
    {\includegraphics[width=.4\linewidth]{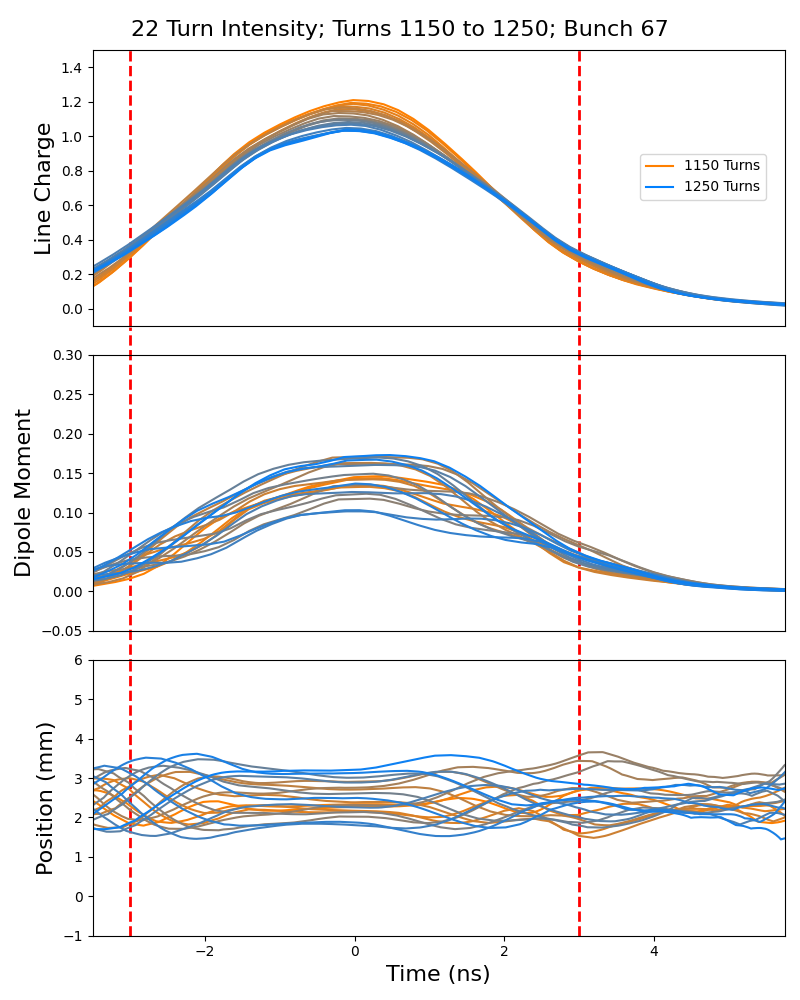}}\qquad
  \subcaptionbox{Stabilization of motion during turns 1250-1400, when the beam leaves transition-crossing regime. \label{fig:Instability_4}}
    {\includegraphics[width=.4\linewidth]{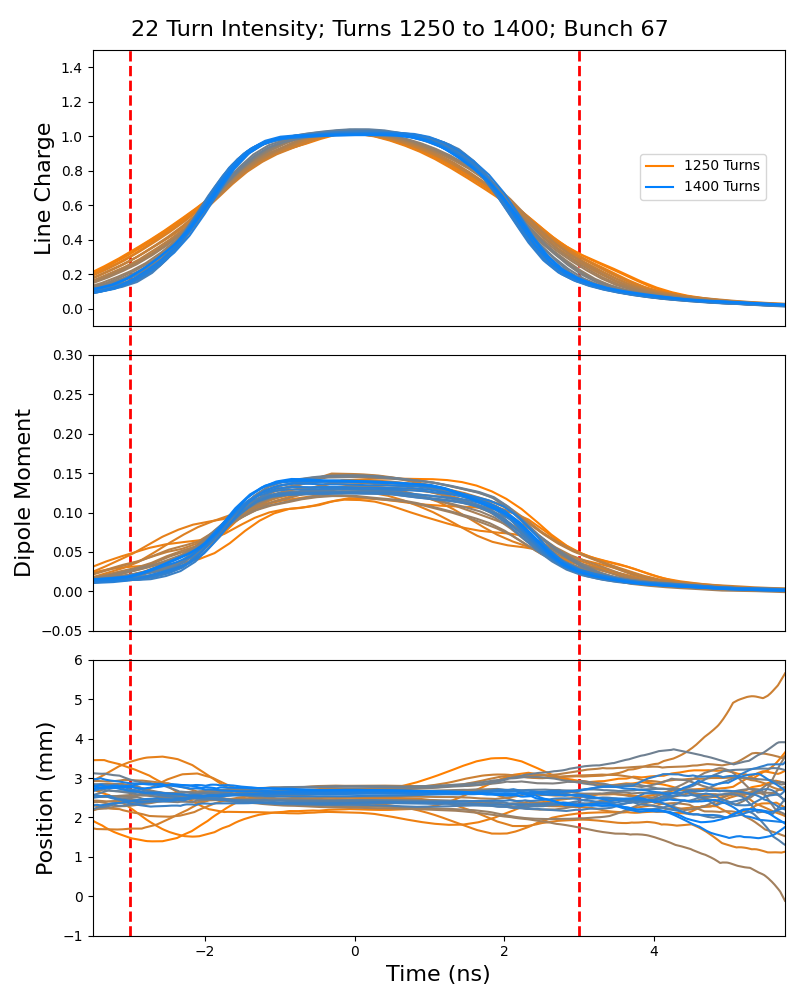}}
    \caption{Line charge, dipole moment, and position along the bunch for three different times in the instability sequence. Transition take places around turn 1140. Position calculations at the edges of the figure are unreliable due to the small line charge.
    \label{fig:Instability_Sequence}}
\end{figure}

Figure~\ref{fig:Instability_Sequence} shows the line-charge, dipole moment, and position across the length of a single bunch undergoing a transverse beam instability over time near transition. In Figure~\ref{fig:Instability_1} with synchrotron motion, we see transverse stability with possibly a small betatron oscillation. In Figure~\ref{fig:Instability_2} leading up to transition, we see exponential growth in the transverse oscillation, with the greatest amplitude of oscillation in the tail. In Figure~\ref{fig:Instability_3} after transition, we see loss across the bunch, but dominated by tail-loss, while the oscillation amplitude may be constrained by aperture restriction.  In Figure~\ref{fig:Instability_4} after beam loss, the oscillation amplitude decoheres and the beam reaches a new equilibrium with lower intensity as synchrotron motion and chromaticity are restored.

The instability observed is consistent with a convective instability that becomes absolutely unstable beyond an intensity threshold. At low intensity, transverse motion of the head and tail is symmetric; as intensity increases large tail amplitudes are observed below the instability threshold. This differs from the form for TMCI without space-charge~\cite{Metral21}, where the instability threshold is below the observation of larger tail amplitudes.

There is ongoing work into the nature of this instability in the Booster, including numerical, analytical, and experimental approaches. Numerically, a multi-loop potential well model~\cite{Balcewicz23} is used to explore mode coupling instabilities with strong space-charge as a function of space-charge and wake parameters. Analytically, the dependence of convective instabilities on chromaticity is being explored. Experimentally, a detailed analysis of synchrotron sidebands of the betatron frequency is visualized over time and over the length of the bunch.

\section{Conclusions \& Future Work}

The performance of the Fermilab Booster plays a critical role in the ambitious DUNE/LBNF long-baseline neutrino physics program, and a more comprehensive understanding of the Booster can provide valuable insight for developing the next generation of intensity-frontier machines. We have presented results and analysis for the 2023 International Fermilab Booster Studies, but only the beginning of an ongoing comprehensive investigation on high-intensity beam dynamics.

In this work we develop procedures for optimizing the 400~MeV beam study mode of the Booster and a new injection mismatch procedure for creating low-brightness beams. The methods of sextupole resonance cancellation and the role of linear coupling resonance in the Booster are confirmed. A MADX-SC simulation clearly demonstrates the role of the ${2Q_{y}=13}$ resonance in the rapid emittance blow-up at injection. Next, we must benchmark the MADX-SC adaptive space-charge model against a PIC code~\cite{Priv_Comm_Person_MADX-SC-IMPACT} to study the Fermilab Booster at its highest intensities. This is a challenging case, where the Laslett tune spread spans the ${2Q_{y}=13}$ resonance and as a result the charge-dominated beam is mismatched far from its equilibrium emittance value.

Separately, there appears to be a significant charge-dependent emittance growth and beam loss mechanism, not well understood from operational experience with ramping cycles in the Booster. Although the MADX-SC simulation does not replicate this effect, we believe this is indicates a shortcoming of Booster model, rather than a deficiency of the simulation code. Upcoming work includes incorporating Booster gradient magnet misalignments and improving the Booster gradient magnet model.

The impact of the ${2Q_{y}=13}$ resonance on early losses in high intensity Booster beams in the Booster is very clearly established. Appropriate quadrupole corrector settings to weaken the ${2Q_{y}=13}$ resonance and greatly improve beam performance early in the Booster cycle have been identified. However, that ${2Q_{y}=13}$ correction is also associated with worse beam performance later in the cycle. We believe that simulating and tuning the space-charge induced RDTs is the next step in understanding a more comprehensive optimization of the Fermilab Booster performance in ramping cycles. Understanding the non-equilibrium dynamics between extreme space-charge and the half-integer resonance may be particularly important for the optimization of bunch compression in a future muon collider proton driver~\cite{EldredIPAC24}.

Further work is needed to understand the IPM data and how to calibrate all instrumentation effects. A new emittance measurement method analyzing quadrupole oscillation modes in Booster BPMs, is being developed to independently measure beam quality and cross-calibrate the IPMs.

Finally, transition-crossing in the Booster is an interesting regime for the study of mode coupling instabilities in the presence of strong space-charge. Numerical, analytical, and experimental work is ongoing to understand the classification, features, and mitigation of these instabilities. In the meantime, this beam instability must be mitigated to achieve PIP-II Booster performance goals. For PIP-II era operation of the Booster, a new chromaticity curve will be developed and the possibility of a resonant $\gamma_{T}$-jump will be investigated.

\acknowledgments

We'd like to thank Foteini Asvesta and Tirsi Prebibaj for traveling from CERN and providing many helpful comments on resonance analysis. We'd also like to thank Alexey Burov for his insightful commentary analyzing instabilities with strong space-charge. Thank you to Kent Triplett, John Kuharik, and Salah Chaurize for establishing 400-MeV mode and assisting with operations. Cheng-Yang Tan and Chandra Bhat for assisting with study coordination and providing general feedback. Thanks to Cheng-yang Tan and John Johnstone for many helpful comments on the last several years of 2$Q_{y}$ correction work.

This work was supported through grant DE-SC0020379 with the United States Department of Energy.


\bibliographystyle{JHEP}
\bibliography{biblio.bib}






\end{document}